\documentclass[12pt,a4paper]{article}
\usepackage{jheppub}
\usepackage[T1]{fontenc}
\usepackage{slashed}
\usepackage{caption}
\usepackage{subcaption} 
\usepackage{float}
\usepackage{empheq}%

\usepackage{comment}
\usepackage{xcolor}
\usepackage{color,soul}
\usepackage{bbold}
\usepackage{graphicx}

\usepackage{feynmp}

\everymath{\displaystyle}

\DeclareMathOperator*{\res}{Res}
\def\[{\left[}
\def\]{\right]}
\def\({\left(}
\def\){\right)}

\preprint{}

\title{Double winding condensate CFT}

 \author[]{Indranil
  Halder\note{ihalder@g.harvard.edu} and}  \author[]{Daniel L. Jafferis\note{jafferis@g.harvard.edu}}

\affiliation{Jefferson Physical Laboratory, Harvard University, Cambridge, MA 02138, USA}

\abstract{ We conjecture a two-dimensional conformal field theory built out of a linear dilaton and a compact $\beta\gamma$ system deformed by winding condensates on each of the compact cycles. In particular, explicit expressions of the residues of the correlation functions are provided. As a worldsheet theory, it describes a stringy black hole in AdS$_3$ (supported by the Kalb-Ramond flux) at the Hawking-Page temperature. It is expected to be connected to the Horowitz-Polchinski-like solution near the Hagedorn temperature of thermal AdS$_3$ and Euclidean BTZ blackhole as we vary the temperature.}

\begin{document}

\maketitle

\section{Introduction}

The microscopic origin of the Bekenstein-Hawking entropy of black holes has been one of the motivating questions in quantum gravity for the past half a century. For extremal BPS black holes, exact matches with the counting of D-brane microstates due to Strominger and Vafa \cite{Strominger:1996sh} is one of the major successes of string theory (around the same time Sen showed the entropy of a `small' extremal black hole up to a numerical factor is same as that of fundamental strings \cite{Sen:1995in, Sen:1997is}.\footnote{A blackhole whose curvature at the horizon of the strings frame metric (without any stringy corrections) is of order strings scale at strings scale temperature is called a `small' blackhole. For a small blackhole stringy corrections extend up to the `stretched horizon'. } The numerical factor, capturing stringy corrections in the near horizon region, was fixed later using the attractor mechanism in \cite{ Dabholkar:2004yr, Dabholkar:2004dq}). The most straightforward calculation of D-brane degeneracies is at weak coupling, and the matching relies on the constancy of the BPS index as one dial to the black hole regime at strong coupling. In some cases, at least a subset of the states can be understood directly at strong coupling in the form of horizonless BPS supergravity solutions \cite{Lunin:2001jy, Lunin:2002iz, Kanitscheider:2007wq, Rychkov:2005ji} (see \cite{Bena:2022rna} for a recent review).

Moreover, the existence of a single-gapped BPS state in the ${\cal N}=4$ super-Liouville quantum mechanics obtained by canonical quantization of the 2d super Jackiw-Teitelboim effective theory description of near extremal black holes \cite{Heydeman:2020hhw, Lin:2022zxd}, suggests a 1-1 correspondence of horizonless BPS configurations and species of effective BPS end of the world branes in the 2d description. Each of these would then lead to an associated contribution to the BPS index. Of course, many of these configurations would be expected to involve D-branes and stringy features, so their full description at strong coupling is complicated. 
The situation away from extremality, however, appears to be conceptually much more difficult. This is because `bags of gold' with spatially large regions behind the horizon already overcount the Bekenstein-Hawking entropy even restricting to long wavelength gravity configurations (this can be seen easily in the description with the end of the world brane mentioned above \cite{Gao:2021uro}). This overcounting is the direct origin of the black hole information problem, as the canonical evolution of an evaporating black hole accesses these too numerous states \cite{Penington:2019kki, Balasubramanian:2022fiy, Akers:2022qdl}. 
One possibility is that most of these states become null states due to exponentially fine-tuned non-perturbative corrections to the inner product on Hilbert space \cite{Saad:2019lba}, as suggested by Euclidean replica wormhole calculations. It is of great interest to understand whether UV complete theories like string theory provide an explanation (or alternative) for that fine-tuning, perhaps involving stringy dualities.

Horowitz and Polchinski made the following observation \cite{Horowitz:1996nw} (for earlier work on this see \cite{Bowick:1985af, Susskind:1993ws, Sen:1995in}): consider free bosonic string states in flat-space at an excitation level $N$, where the entropy and mass in strings units both are of order $N^{1/2}$. A Schwarzschild black hole of the same mass and whose curvature at the horizon (ignoring stringy corrections) in string frame metric is of order string scale, i.e, $g_s\sim N^{-1/4}$, also has Bekenstein-Hawking entropy of order $N^{1/2}$ (the observation was extended to include momentum and winding charges in presence of a compact dimension). Thus although the masses of these non-BPS states depend on the string coupling, one might hope to adiabatically track them and match their number at the correspondence point. 

At first sight, this sounds puzzling because the size of a free strings state is order $N^{1/4}$ in strings units, and as we gradually increase the string coupling $g_s$ the Schwarzschild radius of a given string state increases but for $g_s\sim N^{-1/4}$ it is only order string scale. This is resolved by noting that one needs to take into account the change in the size of string states as we increase $g_s$ and it has been shown precisely by Horowitz and Polchinski that, for instance in four spacetime dimensions, indeed for $g_s\sim N^{-1/4}$ the size of a string state at level $N$ is of order the string scale \cite{Horowitz:1997jc}.\footnote{In absence of momentum and winding charges, the temperature of the Horowitz-Polchinski solution \cite{Horowitz:1997jc}  of the self-gravitating string is slightly below the Hagedorn temperature and as we approach the Hagedorn temperature $g_s$ corrections become important. When the charges are non-zero it is possible to decrease the temperature but as we approach too close to the BPS limit again $g_s$ corrections become non-negligible \cite{Chen:2021dsw}. Note that on the contrary in Sen's discussion \cite{Sen:1995in, Sen:1997is}, $g_s$ is kept fixed while one takes the zero temperature limit.} This result was obtained by mapping the discussion of the microcanonical ensemble presented here to the canonical ensemble near the Hagedorn temperature of flat space where the effective contribution to string thermodynamics comes from a highly excited string state. In target space, the highly excited string state is described by a tachyon in the Euclidean time circle carrying non-trivial winding. For a discussion of various aspects of the solution see \cite{Damour:1999aw, Khuri:1999ez, Kawamoto:2015zha, Martinec:2014gka, Balthazar:2022szl, Balthazar:2022hno, Mertens:2015ola}). It is of great interest to understand the accurate description of these states in the strong coupling regime, to determine their relation to the `bag of gold' geometries.

Kutasov suggested a beautiful picture of the two-dimensional (uncharged, i.e., no additional momentum or winding in any internal circle) black hole in flat space with a linear dilaton \cite{Kutasov:2005rr, Giveon:2006pr} (see also \cite{Adams:2005rb, McGreevy:2005ci, Horowitz:2006mr, Silverstein:2006tm}). He showed the Euclidean black hole has a winding tachyon around the time direction that lives close to the Euclidean horizon 
(in the probe approximation) where the local temperature is that of the Hagedorn temperature of flat space (without the black hole) \cite{Atick:1988si}. He argued that as the asymptotic temperature of the black hole increases to the Hagedorn temperature of the `empty' flat space, the tachyon fills all of the space.\footnote{This picture might also hold true for black holes in large dimensions \cite{Chen:2021emg}.}

For momentum and winding charges near extremality of the would-be stringy black hole geometry of Sen \cite{ Dabholkar:2004yr, Dabholkar:2004dq, Sen:2004dp, Hubeny:2004ji} in heterotic string theory, Chen-Maldacena-Witten \cite{Chen:2021dsw} recently found a well-controlled horizon-less solution of the target space equations of motion containing a winding condensate in the Euclidean time direction. 
 A remarkable fact of the solution given in \cite{Chen:2021dsw} is that 
it is closely related to the uncharged version of the Horowitz-Polchinski solution \cite{Horowitz:1997jc}  near the Hagedorn temperature of empty flat space and it reproduces the black hole entropy including the correct numerical factor. As we move away from the extremality (keeping momentum and winding charges fixed) the temperature of the uncharged Horowitz-Polchinski solution decreases and we need to take stringy corrections into account.\footnote{Whereas far away from extremality the black hole solution considered by Sen will be under control.}  Therefore, in the light of the picture laid out by Kutasov, it is a fascinating open question to understand the stringy features of the uncharged Horowitz-Polchinski solution away from the  Hagedorn temperature. 

A major challenge in this program is the lack of knowledge of the exact worldsheet CFT of the stringy winding condensate in the black hole regime. In that light, the case of AdS$_3$ supported only by Kalb-Ramond flux appears to be a good candidate, since the worldsheet theory is describable in the R-NS formalism as the integrable $SL(2,R)$  WZW model at level $k$ (see \cite{Teschner:1997ft, Teschner:1999ug, Giveon:1998ns, Kutasov:1999xu, Maldacena:2000hw, Maldacena:2000kv, Maldacena:2001km, Ribault:2005wp, Dei:2021yom, Dei:2021xgh} and \cite{Gaberdiel:2018rqv, Eberhardt:2018ouy, Eberhardt:2019ywk, Dei:2020zui, Gaberdiel:2020ycd, Martinec:2021vpk, Balthazar:2021xeh, Martinec:2023zha} for conventional sigma model approach). 
For our purpose, we find it is more appropriate to use a dual description of AdS$_3$ in terms of a  winding condensate in a spatial circle around a free theory as developed in \cite{Jafferis:2021ywg, Halder} (also see \cite{Berkooz:2007fe}).
The target space effective theory with a Horowitz-Polchinski-like winding condensate on the temporal circle near the Hagedorn transition point is discussed in \cite{Urbach:2023npi} (which is the analog of the uncharged solution considered above).\footnote{Similar solution involving space winding tachyon exists around Euclidean BTZ blackhole at low enough temperature  \cite{Horowitz:2005vp, Ross:2005ms, Rangamani:2007fz}.} In this paper, we investigate the intriguing possibility that there is an integrable dual description of the worldsheet CFT deep into the stringy `small' black hole regime. 
In particular, at the self-dual Hawking-Page temperature, we construct a deformation of the free theory by a double condensate of winding operators around both the spatial and temporal circles.

From the worldsheet perspective, the Euclidean `string star'/Horowitz-Polchinski-like solution near the Hagedorn temperature would be constructed by analyzing a weakly irrelevant deformation of thermal AdS$_3$ by a time winding operator and finding a zero of the beta function equations including other light operators, truncated to second order in \cite{Nick}. Thermal AdS$_3$ is the periodic quotient of the Euclidean target $SL(2,C)/ SU(2)$ coset along the temporal circle. Far from the Hagedorn temperature, it would seem difficult to find a precise description of the resulting CFT.

Starting from thermal AdS$_3$, the stringy worldsheet theory constructed in this paper is expected to be a finite deformation, associated with a distinct, isolated black solution to the target space equations of motion with the same asymptotics (up to subtleties involving a locally flat Kalb-Ramond field on the boundary torus). A finite deformation here refers to the fact that there is not an exactly marginal family of CFTs connecting these two theories while keeping fixed the temperature. 
Tuning close to the Hagedorn temperature makes this deformation parametrically small, and controllable in conformal perturbation theory. However, in general, it is a non-perturbative deformation of the conformal field theory. 

We will show that at the Hawking-Page temperature, there is a candidate theory with a simple description in terms of a combination of spatial and temporal winding condensates. This is based on the duality between thermal AdS$_3$ and a CFT built on a linear dilaton and compact $\beta\gamma$ system deformed by a marginal winding operator. We check some basic consistency conditions, including the vanishing of one-point functions, as required by conformality. This is not sufficient to fix the relative coefficient of the winding condensates, which from the target space perspective must be determined by regularity in the interior. We conjecture that when the coefficients are equal, there is a non-singular CFT with target space AdS$_3$ asymptotics.

As usual in AdS, there are two possible fall-offs for each mode in the asymptotic region, and regularity in the interior relates their coefficients in the Euclidean theory. In the highly stringy $2<k<4$ regime,  both winding operators and their reflected waves decay at asymptotic infinity, thus, given our results, the theory with both condensates presumably exists for any value of the relative coefficient. Changing this parameter is still associated with a non-normalizable vertex operator. Such a scenario is analogous to the family of worldsheet CFTs associated with sine-Liouville theory with a temperature different from the cigar coset temperature, in the $2<k<4$ regime. For $k\geq  4$, to demonstrate the existence of a well-defined CFT with asymptotically AdS target space, one would need to check that the reflection amplitude of the condensates exactly vanishes in the two winding deformed theory. We leave this important problem for future work.

The paper is organized as follows. In section \ref{2}, we carefully review the description of \cite{Jafferis:2021ywg, Halder} and point out a subtlety in the definition of conserved translation current in non-trivial winding sectors related to the discussion in \cite{Rangamani:2007fz}.
In section \ref{3}, marginal vertex operators in the free theory carrying winding in space and time circles are discussed. It is shown that precisely at the Hawking-Page temperature the spatial winding condensate that builds AdS, and the Euclidean time winding condensate that preserves U(1)$\times$U(1) translational symmetry, are both marginal with the same radial profile. Based on this observation, in section \ref{4} we systematically study the free theory deformed by these double winding condensate \textit{exactly} using the technique of \cite{Halder}. For the simplest set of vertex operators in the theory we provide an integral representation for the residue of arbitrary correlation functions. Next, we turn to show that the integral representation for the two and three-point function has the expected properties of a conformal field theory. To this end, we point out a connection between the integrals mentioned above with the theory of hypergeometric integrals. Utilizing this connection, we show that the integral representation for the residue of the three-point function has a non-empty domain of convergence. As a result, formal conformal invariance properties are maintained when we analytically continue away from this domain. We expect this to be true for higher point correlation functions as well. Furthermore, we show that one point functions of non-trivial operators vanish in the theory as is necessary for it to be conformal. Our considerations do not allow us to fix the relative coefficient between the two winding condensates, however, we believe additional considerations of `regularity in the interior' (vanishing of the self-reflection coefficients) would fix it to a specific value. In section \ref{5} we discuss interesting future directions.

\section{Euclidean AdS$_3$}\label{2}

In this section, we will consider bosonic string theory in AdS$_3 \times$X, where X is a compact manifold. 
When the AdS length ($\sqrt{k}$) is large in strings units, the sigma model description of the worldsheet CFT in the Poincare patch takes the following form
\begin{equation}\label{sigma_model}
	\begin{aligned}
		& S_{cl}=\frac{k}{2 \pi}\int 2 d^2 \sigma \(\partial \hat{\phi} \bar{\partial} \hat{\phi} +\bar{\partial} \hat{\Gamma} \partial \bar{\hat{\Gamma}}e^{2\hat{\phi}} \)\\
	\end{aligned}
\end{equation}
It is well known that a natural completion to all orders in $\alpha'$ is obtained by considering the WZW model based on group SL(2,C)$_k$/SU(2). However for the purpose of this paper we will work in the global patch and use the dual description of the worldsheet CFT developed in \cite{Jafferis:2021ywg, Halder}.  We turn to review this formalism briefly and discuss new subtleties that show up in the definition of conserved U(1) current on the worldsheet.

To get the worldsheet in global co-ordinates (at large $k$) we make the following change of coordinates
\begin{equation}
	\begin{aligned}
		& \hat{\phi}:=-\hat{\xi}+\log \cosh \hat{r},~~ \hat{\Gamma}:=\tanh \hat{r} \ e^{\hat{\xi}+i \hat{\theta}},~~ \hat{\bar{\Gamma}}:=\tanh \hat{r} \ e^{\hat{\xi}-i \hat{\theta}}, ~~ \hat{\theta}\sim \hat{\theta}+2\pi
	\end{aligned}
\end{equation}
The action takes the following form 
\begin{equation}\label{globalAdS}
	\begin{aligned}
		& S_{cl}=\frac{k}{2 \pi}\int 2 d^2 \sigma \(\partial \hat{r} \bar{\partial} \hat{r}+\partial \hat{\xi} \bar{\partial} \hat{\xi} +\sinh^2  \hat{r} \  \partial (\hat{\xi} -i \hat{\theta})\bar{\partial} (\hat{\xi} +i \hat{\theta})+ i \tanh \hat{r} (\partial \hat{r} \bar{\partial} \hat{\theta}-\partial \hat{\theta} \bar{\partial} \hat{r})\)\\
	\end{aligned}
\end{equation}
with holomorphic conserved current\footnote{This matches with \cite{Maldacena:2000hw} up to an overall factor of $i$ with the replacement $\gamma=2iu, \bar{\gamma}=2i v$.}
\begin{equation}\label{CurrentGlobal}
	J^3=-\frac{k}{2}(\partial \gamma +\cosh(2 \hat{r})\partial \bar{\gamma})
\end{equation}
Up to total derivatives, the above action can be simplified as follows (for the rest of this sub-section we will not make any changes to the choice of total derivatives made here)\footnote{Ignoring total derivatives might be subtle sometimes. For example, if two classical backgrounds differ by a flat $B$ field with non-trivial flux on the torus at the boundary of AdS$_3$, we will get similar additional total derivatives. However, this means that the backgrounds differ by a non-normalizable deformation on the bulk of  AdS$_3$. We will assume such subtleties are not important in the proposed dual description. }
\begin{equation}\label{globalAdS2}
	\begin{aligned}
		 S_{cl}=& \frac{k}{4 \pi}\int d^2 \sigma (4 \ \partial \hat{r} \bar{\partial} \hat{r}+\partial \gamma \bar{\partial}\gamma +\partial \bar{\gamma} \bar{\partial}\bar{\gamma}+2(\cosh 2\hat{r}+c) \  \partial \bar{\gamma} \bar{\partial}\gamma -2c \ \partial \bar{\gamma} \bar{\partial}\gamma)\\
	\end{aligned}
\end{equation}
Here 
\begin{equation}
	\begin{aligned}
		&\gamma=\hat{\xi}+i \hat{\theta}, ~~~~~~ \bar{\gamma}:=\hat{\xi}-i \hat{\theta}
	\end{aligned}
\end{equation}
We want to go to the first-order formalism where it is easy to drop the explicit interaction term. There are various inequivalent ways of doing so, i.e., choosing the values of $c$. At large $k$,
\begin{equation}\label{action}
	\begin{aligned}
		& S_{cl}=\frac{1}{2 \pi}\int 2 d^2 \sigma \(k \ \partial \hat{r} \bar{\partial} \hat{r}+(\chi+\frac{k}{4}\partial \gamma-\frac{k}{4}c\partial \bar{\gamma} )\bar{\partial}\gamma +(\bar{\chi}+\frac{k}{4}\bar{\partial} \bar{\gamma}-\frac{k}{4}c\bar{\partial} \gamma)\partial \bar{\gamma} -\frac{1}{\frac{k}{2}(\cosh 2\hat{r}+c)} \chi \bar{\chi} \)\\
	\end{aligned}
\end{equation}
The equivalence at the classical level is easily seen by the following equations of motion
\begin{equation}\label{eom}
	\chi=\frac{k}{2}(\cosh 2\hat{r}+c)  \ \partial \bar{\gamma}, ~~ \bar{\chi}=\frac{k}{2}(\cosh 2\hat{r}+c) \ \bar{\partial} \gamma
\end{equation}
It is convenient to consider following set of co-ordinates ($\beta$, $\bar{\beta}$ are complex conjugate of each other as long as $a$ is real)
\begin{equation}
	\begin{aligned}
		& \varphi=-\sqrt{k} \ \hat{r},~~ \beta=\chi+\frac{k}{4}\partial \gamma-\frac{k}{4}c\partial \bar{\gamma}, ~~ \bar{\beta}=\bar{\chi}+\frac{k}{4}\bar{\partial} \bar{\gamma}-\frac{k}{4}c\bar{\partial} \gamma
	\end{aligned}
\end{equation}
Near the boundary of $AdS$ the action (\ref{action}) reduces to the following free theory (valid for all $k$)
\begin{equation}
	\tilde{S}=\frac{1}{2 \pi}\int 2 d^2 \sigma (\partial \varphi \bar{\partial} \varphi+\frac{Q}{4}\varphi R +\beta \bar{\partial} \gamma +\bar{\beta} \partial \bar{\gamma}), ~~ Q=\frac{1}{\sqrt{k-2}}
\end{equation} 
Relevant OPE are
\begin{equation}\label{OPE}
	\begin{aligned}
		\beta(z)\gamma(0)\sim -\frac{1}{z}, ~~~ \varphi(z, \bar{z})\varphi(0,0)\sim -\frac{1}{2}(\ln z+\ln \bar{z})
	\end{aligned}
\end{equation}
The holomorphic stress tensor is given by
\begin{equation}
	\begin{aligned}
		& T=- (\partial \varphi)^2+Q\partial^2 \varphi-\beta \partial \gamma\\
	\end{aligned}
\end{equation}
The conserved holomorphic $U(1)$ current  is given by (it is just the current in (\ref{CurrentGlobal}) written in new variables using equations of motion in the interacting theory (\ref{eom}))
\begin{equation}\label{theCurrent}
	\begin{aligned}
		& J^3=-\beta -\frac{k}{4}\partial \gamma+\frac{k}{4}c\partial \bar{\gamma}+c'\partial(\gamma +\bar{\gamma})
	\end{aligned}
\end{equation}
Here $c,c'$ are two undetermined constants (eventually these will be fixed to $0,k/4$ respectively). 
The currents corresponding to $\hat{\xi}, \hat{\theta}$ are $J^3_{\hat{\xi}}=J^3+\bar{J}^3, J^3_{\hat{\theta}}=J^3-\bar{J}^3$ respectively.  Note that the apparent total derivatives $\partial \gamma,\partial \bar{\gamma} $ can be ignored only when we are looking at the action of $J^3_0$ on operators that do not carry winding charge in $\hat{\xi}, \hat{\theta}$; in other words when both $\gamma, \bar{\gamma}$ are single-valued. For example, when we act on operators which carry non-trivial winding in $\hat{\theta}$ direction but zero winding in $\hat{\xi}$ direction we can only ignore the particular combination $\partial(\gamma+\bar{\gamma})$ in the expression for $J^3$ for the purpose of evaluating $J^3_0$. 

First, we focus on vertex operators that carry winding only around the space circle. These take the following form\footnote{Here $\int_{}^{(z,\bar{z})}$ instructs us to only keep the value at the upper limit of the integration. A more precise way of saying it is that as long as we are looking at the correlation function of ``allowed'' local operators, neither the choice of contour nor the lower limit of integration will play any role. For a discussion of this property in the toric sigma model see \cite{Frenkel:2005ku}.}
\begin{equation}\label{vo}
	\begin{aligned}
		\tilde{V}^n_{\alpha,a,\bar{a}}=\frac{1}{\pi} \ e^{a \gamma+\bar{a} \bar{\gamma}} \ e^{n(\int_{}^{(z,\bar{z})}\beta dz'+ \int_{}^{(z,\bar{z})}\bar{\beta} d\bar{z}')} \ e^{2\alpha \varphi}
	\end{aligned}
\end{equation}

The charges under $L_0, J^3_0$ of the operator (as determined by the OPE with $T, J^3$) take the following form
\begin{equation}
	h'=\alpha(Q-\alpha)-na, ~~ m'=a+\frac{kn}{4}+\frac{kn}{4}c
\end{equation}
Now we are in a situation to fix the value of $c$ that corresponds to the empty AdS background. To this end we compare the expression above, with the one in \cite{Maldacena:2000hw} for a vertex operator which corresponds to the SL(2,R) Casimir labeled by $j$, and the $J_0^3$ eigenvalue $m$ before spectral flow by $\nu$ units\footnote{This matches with \cite{Maldacena:2000hw} with the identification of $m (\text{here})= -m(\text{there})$. }
\begin{equation}
	h'=-\frac{j(j+1)}{k-2}+\nu(m-\frac{k}{4}\nu),~~~ m'=m-\frac{k}{2}\nu
\end{equation}
This determines
\begin{equation}
	c=0
\end{equation}
with the following map
\begin{equation}
	\begin{aligned}
		& \alpha=-Qj,~~ 
		& a=m'+\frac{k\nu}{4}, ~~
		& n=-\nu
	\end{aligned}
\end{equation}
The number $c'$ will be determined in the next section.

The geometric meaning of the spectral flow becomes clear from the following OPE
\begin{equation}
	\begin{aligned}
		& \gamma(z, \bar{z}) e^{n \int_{}^{(0,0)}\beta dz'} \sim -n \log (z) \ e^{n \int_{}^{(0,0)}\beta dz'} \\
		& \bar{\gamma}(z, \bar{z}) e^{n \int_{}^{(0,0)}\bar{\beta} d\bar{z}'} \sim -n \log (\bar{z}) \ e^{n \int_{}^{(0,0)}\bar{\beta} d\bar{z}'}
	\end{aligned}
\end{equation}
This implies if we rotate $\hat{\theta}=(\gamma-\bar{\gamma})/(2i)$ around the insertion of the operator  $e^{-n(\int_{}^{(0,0)}\beta dz'+ \int_{}^{(0,0)}\bar{\beta} d\bar{z}')}$ on the worldsheet we get $\hat{\theta}\to \hat{\theta}+2\pi n$, i.e., the operator  $e^{-n(\int_{}^{(0,0)}\beta dz'+ \int_{}^{(0,0)}\bar{\beta} d\bar{z}')}$ has winding $n$ around the $\hat{\theta}$ circle. Similar arguments show that this operator has no winding around the $\hat{\xi}$ circle. If we compactify the time direction as well as $$\hat{\xi} \sim \hat{\xi}+R$$ then $e^{i w \frac{R}{2 \pi}(\int_{}^{(0,0)}\beta dz'-\int_{}^{(0,0)}\bar{\beta} d\bar{z}')}$ carries $w$ units of winding around the  $\hat{\xi}$ circle and carries no winding around the $\hat{\theta}$ circle.

For the purpose of discussing thermal AdS$_3$, we need more general vertex operators compared to the ones presented in (\ref{vo}), which are given by
\begin{equation}\label{vertexOp}
	\begin{aligned}
		\tilde{V}^{s,\bar{s}}_{\alpha,a,\bar{a}}=N^{s,\bar{s}}_{\alpha,a,\bar{a}}\ e^{a \gamma+\bar{a} \gamma} \ e^{s\int_{}^{(z,\bar{z})}\beta dz'+\bar{s} \int_{}^{(z,\bar{z})}\bar{\beta} d\bar{z}'} \ e^{2\alpha \varphi}
	\end{aligned}
\end{equation}
Winding around the $\hat{\theta}$ circle is given by
\begin{equation}
	n=-\frac{s+\bar{s}}{2}
\end{equation}
Winding around the $\hat{\xi}$ circle is given by
\begin{equation}
	w=-\frac{\pi }{R}(s-\bar{s})
\end{equation}

\section{Two winding condensates}\label{3}

We have two distinguished vertex operators which conserve both $J^3_{0,\hat{\xi}}, J^3_{0,\hat{\theta}}$, are marginal on the worldsheet and carry $\pm 1$ winding respectively in the $\hat{\theta}$ direction (and no winding in the $\hat{\xi}$ direction).
\begin{subequations}\label{SpaceWinding}
\begin{empheq}{align} 
  & V^{\pm}=\ e^{\pm \frac{k}{4}(\gamma+\bar{\gamma})} \ e^{\mp  (\int_{}^{(z,\bar{z})}\beta dz'+ \int_{}^{(z,\bar{z})}\bar{\beta} d\bar{z}')} \ e^{2b\varphi}, ~ b=\frac{1}{2}\sqrt{k-2}
\end{empheq}
\end{subequations}
The dual description of AdS$_3$ sigma model action takes the following form \cite{Halder}
\begin{equation}\label{TAdS}
	S_{TAdS}=\frac{1}{2 \pi}\int 2 d^2 \sigma (\partial \varphi \bar{\partial} \varphi+\frac{Q}{4}\varphi R +\beta \bar{\partial} \gamma +\bar{\beta} \partial \bar{\gamma} +\frac{\pi\mu}{\alpha_+^2} (V^{+}+ V^{-}))\\
\end{equation}
The measure for path integration is the usual Liouville measure for $\varphi$ and the standard measure for the $\beta\gamma$ system. The interaction coupling $\mu$ is given by
\begin{subequations}
\begin{empheq}{align}\label{duality_map}
	& ~~b:=\frac{b'}{2}, ~	\mu:=b'^2 \alpha_+^2\(\frac{\mu'\gamma(b'^2)}{\pi }\)^{1/2}\\
	 & b':=\frac{1}{b''}, ~~ \pi \mu'\gamma(b'^2):=(\pi \mu''\gamma(b''^2))^{1/(b''^2)}\\
	 & b''=\frac{1}{\sqrt{k-2}}, ~~ \mu''=\frac{b''^2}{\pi^2}
\end{empheq}
\end{subequations}
The first of these equations defines the Liouville parameter $(b',\mu')$ which is then related to $(b'',\mu'')$ through the Liouville duality. The last of these equations is a well-known map relating sigma model correlation functions and those of a Liouville theory \cite{Ribault:2005wp}.

There is another set of natural vertex operators which conserve $J^3_{0,\hat{\xi}}$, are marginal and carry $\pm 1$ winding in $\hat{\xi}$ direction (and no winding in $\hat{\theta}$ direction)
\begin{subequations}\label{TimeWinding}
\begin{empheq}{align} 
  & W^{\pm}=\ e^{\pm i \frac{R}{2\pi} \frac{k}{4}(\gamma-\bar{\gamma})} \ e^{\pm i \frac{R}{2\pi}  (\int_{}^{(z,\bar{z})}\beta dz'- \int_{}^{(z,\bar{z})}\bar{\beta} d\bar{z}')} \ e^{2 (\frac{Q}{2}+iP) \varphi}\\
  & ~~~~~~~~~~~P^2+\(\frac{Q}{2}\)^2+\frac{k}{4}\(\frac{R}{2\pi} \)^2=1
\end{empheq}
\end{subequations}
These operators also conserve $J^3_{0,\hat{\theta}}$ provided we choose $c'=k/4$, i.e., \footnote{Similar total derivative improvement term in the expression of current is suggested around BTZ black hole - see for example eq. (3.15) of \cite{Rangamani:2007fz}. }
\begin{equation}
	\begin{aligned}
		& J^3=-\beta -\frac{k}{4}\partial \gamma+\frac{k}{4}\partial(\gamma +\bar{\gamma})=-\beta +\frac{k}{4}\partial \gamma+\frac{k}{4}\partial(\bar{\gamma}-\gamma)
	\end{aligned}
\end{equation}
Then we have
\begin{equation}
	\begin{aligned}
		m'=a+\frac{k}{4}n \text{ for }w=0 , = a-\frac{k}{4}n \text{ for }n=0 
	\end{aligned}
\end{equation}

At this point, we pause for a few important observations. We note that the time winding operator (\ref{TimeWinding}) is tachyonic, i.e., violets Breitenlohner-Freedman
(BF) bound, 
for temperatures higher than the Hagedorn temperature \cite{Berkooz:2007fe}
\begin{equation}
	R=R_H=2 \pi \( \frac{1}{k}\(4-\frac{1}{k-2}\)\)^{1/2} 
\end{equation}
As a result, these vertex operators represent the stringy state that goes massless at the Hagedorn temperature and causes a divergence in the thermal partition function. At a generic temperature, if we add $W^{\pm}$ to the action of thermal AdS$_3$, we are expected to get a non-trivial RG flow on the worldsheet. Slightly below Hagedorn temperature, a perturbative solution for the new CFT is possible when additional deformations on the worldsheet are turned on along with $W^\pm$. This solution is the worldsheet analog of the Horowitz-Polchinski-like solution in the target space. Something remarkable happens at the Hawking-Page temperature
\begin{equation}
	R=R_{HP}=2\pi
\end{equation}
At this temperature, the radial dependence of $V^\pm$ and $W^\pm$ becomes the same. Conceptually this might have been expected because at this temperature both the time and space circles are on an equal footing. As a result at this special temperature, it is possible to study the theory below
\begin{equation}\label{StringyBlackhole}
	\begin{aligned}
		S_{SBH}=S_{TAdS}+\frac{\pi \lambda}{\alpha_+^2}(W^++W^-)
	\end{aligned}
\end{equation}
using the technique of fields \cite{Halder}. It is natural to expect that the theory in (\ref{StringyBlackhole}) is a conformal field theory at this temperature. In the next section, we will analyze this possibility in detail.

\section{A CFT with two winding condensates}\label{4}

We would like to compute the following correlation function in the worldsheet CFT given by (\ref{StringyBlackhole}) at Hawking-Page temperature $R=R_{HP}$ with the usual measure on the Liouville field $\varphi$ and $\beta\gamma$ system (see  \cite{Halder}  for more details on normalization of zero modes). More explicitly we are looking at the following worldsheet CFT
\begin{equation}
\begin{aligned}\label{doublewindingaction}
	& S_{SBH}=\frac{1}{2 \pi}\int 2 d^2 \sigma (\partial \varphi \bar{\partial} \varphi+\frac{Q}{4}\varphi R +\beta \bar{\partial} \gamma +\bar{\beta} \partial \bar{\gamma} +\frac{\pi\mu}{\alpha_+^2} (V^{+}+ V^{-})+\frac{\pi \lambda }{\alpha_+^2}(W^++W^-))\\
	& V^{\pm}=\ e^{\pm \frac{k}{4}(\gamma+\bar{\gamma})} \ e^{\mp  (\int_{}^{(z,\bar{z})}\beta dz'+ \int_{}^{(z,\bar{z})}\bar{\beta} d\bar{z}')} \ e^{2b\varphi}\\
	& W^{\pm}=\ e^{\pm i \frac{k}{4}(\gamma-\bar{\gamma})} \ e^{\pm i   (\int_{}^{(z,\bar{z})}\beta dz'- \int_{}^{(z,\bar{z})}\bar{\beta} d\bar{z}')} \ e^{2 b \varphi}
\end{aligned}
\end{equation}
It is easy to check from the OPE that $V^\pm W^\pm$ are mutually local with each other. 

We will restrict ourselves to sphere topology and calculate the correlation function of operators with no winding
\begin{equation}
	\begin{aligned}
		\langle \prod_{a} \tilde{V}^{0,0}_{\alpha,a,\bar{a}}(w_a,\bar{w}_a)  \rangle
	\end{aligned}
\end{equation}
Before discussing the correlation function of these operators we need to check their mutual locality with respect to the winding operators $V^\pm, W^\pm$. The OPE has the following structure  
\begin{equation}
	\begin{aligned}
	&	V^+(z, \bar{z}) \tilde{V}^{0,0}_{\alpha,a,\bar{a}}(0,0) \sim (z)^{-2\alpha b+a} (\bar{z})^{-2\alpha b +\bar{a}} (1+\dots) \\ 
	&	V^-(z, \bar{z}) \tilde{V}^{0,0}_{\alpha,a,\bar{a}}(0,0) \sim (z)^{-2\alpha b-a}(\bar{z})^{-2\alpha b -\bar{a}}(1+\dots)\\
	&	W^+(z, \bar{z}) \tilde{V}^{0,0}_{\alpha,a,\bar{a}}(0,0) \sim (z)^{-2\alpha b-ia} (\bar{z})^{-2\alpha b +i\bar{a}} (1+\dots)\\ 
	&	W^-(z, \bar{z}) \tilde{V}^{0,0}_{\alpha,a,\bar{a}}(0,0) \sim (z)^{-2\alpha b+ia}(\bar{z})^{-2\alpha b -i\bar{a}}(1+\dots)\\
	\end{aligned}
\end{equation}
The vertex operators $\tilde{V}^{0,0}_{\alpha,a=m,\bar{a}=\bar{m}}$ that are mutually local with respect to $V^\pm, W^\pm$ are precisely the ones with $m - \bar{m} \in \mathbb{Z},m +\bar{m} \in i\mathbb{Z} $, corresponding to appropriately quantized momentum around both cycles of the boundary $T^2$. Note that a complete set of operators would additionally include operators carrying non-trivial winding on the space and time circles. In this work, we will only focus on the simplest ones, of the form  $$\tilde{V}^{0,0}_{\alpha,0,0}$$

For the purpose of calculation of correlation involving these operators, we perform the path integral over the Liouville zero mode $\varphi_0$ defined by ($\varphi_0$ denotes the kernel of the scalar Laplacian and $\varphi'(z,\bar{z})$ is the space of functions orthogonal to the kernel) \cite{PhysRevLett.66.2051, Dorn:1994xn, Zamolodchikov:1995aa, Teschner:2001rv}
\begin{equation}
	\varphi(z,\bar{z})=\varphi_0+\varphi'(z,\bar{z})
\end{equation}
using the following identity
\begin{equation}
	\begin{aligned}
	\int_{-\infty}^{+\infty}d\varphi_0 \ e^{2a\varphi_0-\alpha e^{2b\varphi_0}}=\frac{1}{2b}\Gamma\(\frac{a}{b}\)\alpha^{-\frac{a}{b}}
	\end{aligned}
\end{equation}
We are left with the following integral over $\varphi'$ alone
\begin{equation}\label{cor1}
	\begin{aligned}
		& ~~\langle \prod_{a} \tilde{V}^{0,0}_{\alpha_a,0,0}(w_a,\bar{w}_a)  \rangle\\
		=& ~~\frac{\pi}{b}\Gamma\(-s\)\langle \prod_{a} \tilde{V}^{0,0}_{j_a, 0,0} (w_a,\bar{w}_a) (\frac{\mu}{\alpha_+^{2}}\int  d^2 z \ (V^{+}(z,\bar{z})+ V^{-}(z,\bar{z}))\\
		&~~~~~~~~~~~~~~~~~~~~~~~~~~~~~~~~~~~+\frac{\lambda}{\alpha_+^{2}}\int  d^2 z \ (W^{+}(z,\bar{z})+ W^{-}(z,\bar{z})))^{s}\rangle'
	\end{aligned}
\end{equation}
We have defined the following Liouville-like parameters 
\begin{equation}\label{ac}
	s=\frac{Q-\sum_a \alpha_a}{b} =\frac{1+\sum_aj_a}{2b^2}
	\end{equation}
Here $\langle .. \rangle'$ denotes path integration only over $\varphi'$ with  the action $\tilde{S}$ under replacement $\varphi\to\varphi'$ and $V^\pm \to 0, W^\pm \to 0$. Therefore for the path integration only over $\varphi'$ we have additional conservation laws that we mention below.

\textbf{Winding conservation:}
In the free theory winding around both circles are conserved. Since external operators do not carry any winding, the only contribution comes from the terms of the form $(V^+)^{s_1}(V^-)^{s_1}(W^+)^{s_2}(W^-)^{s_2}$. 
We will assume $s_1, s_2$ are integers for the argument presented here. Because of the factor of $\Gamma(-s)$ in (\ref{cor1}) we get a pole for positive integer values of $s=2s_1+2s_2$. We will be interested in the residue at these poles. For odd $s$ the free field analysis would still be valid and suggest that the residue at these poles vanishes, in other words, we have pole only for even integer $s$. We will look at these for the time being.

\textbf{Momentum conservation:}
This gives us the following selection rule (note that we do not get any more constraints on $s_1,s_2$ because the momentum of $W^+,W^-$ are precisely opposite to each other and therefore winding conservation in $\hat{\xi}$ direction and momentum conservation does not give independent constraints.)
\begin{equation}\label{ac2}
	\begin{aligned}
		\sum_a m_a=0, ~~ \sum_a \bar{m}_a=0
	\end{aligned}
\end{equation}
These are automatically satisfied for the vertex operators in question. Therefore unless $\mu \lambda=0$ we have to sum over all the allowed values of $s_1,s_2$ (one effective free parameter). The calculation of residues is much more involved compared to those of $SL(2,R)$ WZW model, given explicitly by
\begin{equation}\label{CorFunGen}
	\begin{aligned}
		&~~\res_{s\to 2s_1+2s_2} \ ~~\langle \prod_{a} \tilde{V}^{0,0}_{\alpha_a,a_a,\bar{a}_a}(w_a,\bar{w}_a)  \rangle\\
		=& ~~\frac{\pi}{b} \frac{(-1)^s}{s!} \sum_{2s_1+2s_2=s} \frac{(2s_1+2s_2)!}{(s_1!)^2 (s_2!)^2} \frac{\mu^{2s_1}}{\alpha_+^{4s_1}} \frac{\lambda^{2s_2}}{\alpha_+^{4s_2}} \ \prod_a N^{0,0}_{\alpha_a,a_a,\bar{a}_a} 
		~~\prod_{a<b} \[ w_{ab}^{-2\alpha_a\alpha_b}  \bar{w}_{ab}^{-2\alpha_a\alpha_b} \] \\&~~~~~~~~~~~~~~~ \int\prod_{i=1}^{s_1}  d^2z_i \int \prod_{i'=1}^{s_1}  d^2z'_{i'} \\
		&~~~~~~~~~~~~~~~~\prod_{i=1}^{s_1}\prod_{a} \[ (z_i-w_a)^{-2\alpha_a b+a_a}  (\bar{z}_i-\bar{w}_a)^{-2\alpha_a b +\bar{a}_a}\]\\
		&~~~~~~~~~~~~~~~~\prod_{i'=1}^{s_1}  \prod_{a} \[ (z'_{i'}-w_a)^{-2\alpha_a b-a_a}  (\bar{z}'_{i'}-\bar{w}_a)^{-2\alpha_a b -\bar{a}_a} \]\\
		&~~~~~~~~~~~~~~~~\prod_{i'<j'}  \[ (z_{i'j'})^{\frac{k}{2}-2b^2}  (\bar{z}_{i'j'})^{\frac{k}{2}-2b^2}\] \ \prod_{i<j}   \[ (z_{ij})^{\frac{k}{2}-2b^2}  (\bar{z}_{ij})^{\frac{k}{2}-2b^2}\]\\
		&~~~~~~~~~~~~~~~~ \prod_{i,j'} \[ (z_{ij'})^{-\frac{k}{2}-2b^2}  (\bar{z}_{ij'})^{-\frac{k}{2}-2b^2}\]\\
	\end{aligned}
\end{equation}
\begin{equation}
	\begin{aligned}
		&~~~~\int\prod_{I=1}^{s_2}  d^2z_I \int \prod_{I'=1}^{s_2}  d^2z'_{I'} \\
		&~~~~\prod_{I=1}^{s_2}\prod_{a} \[ (z_I-w_a)^{-2\alpha_a b-ia_a}  (\bar{z}_I-\bar{w}_a)^{-2\alpha_a b +i\bar{a}_a}\]\\
		&~~~~\prod_{I'=1}^{s_2}  \prod_{a} \[ (z'_{I'}-w_a)^{-2\alpha_a b+ia_a}  (\bar{z}'_{I'}-\bar{w}_a)^{-2\alpha_a b -i\bar{a}_a} \]\\
		&~~~~\prod_{I'<J'}  \[ (z_{I'J'})^{\frac{k}{2}-2b^2}  (\bar{z}_{I'J'})^{\frac{k}{2}-2b^2}\] \ \prod_{I<J}   \[ (z_{IJ})^{\frac{k}{2}-2b^2}  (\bar{z}_{IJ})^{\frac{k}{2}-2b^2}\]\\
		&~~~~ \prod_{I,J'} \[ (z_{IJ'})^{-\frac{k}{2}-2b^2}  (\bar{z}_{IJ'})^{-\frac{k}{2}-2b^2}\]\\
		&~~~\\
		&~~~~ \prod_{i,J} \[ (z_{iJ})^{-2b^2}  (\bar{z}_{iJ})^{-2b^2}\]\\
		&~~~~ \prod_{i',J} \[ (z_{i'J})^{-2b^2}  (\bar{z}_{i'J})^{-2b^2}\]\\
		&~~~~ \prod_{i,J'} \[ (z_{iJ'})^{-2b^2}  (\bar{z}_{iJ'})^{-2b^2}\]\\
		&~~~~ \prod_{i',J'} \[ (z_{i'J'})^{-2b^2}  (\bar{z}_{i'J'})^{-2b^2}\]\\
	\end{aligned}
\end{equation}
Here we use the convention that $V^+,V^-,W^+,W^-$ are at $z_i, z_{i'}, z_I, z_{I'}$ and $z_{i'J}=z'_{i'}-z_J$ etc. Remarkably for $k \to 2$ the coupling between $z_i, z'_{i'}$ and $z_I, z'_{I'}$ vanishes as can be seen from (\ref{cor1}). In this case, we can write the residue of the correlation function in the double winding condensate CFT in terms of a sum over the product of residues in single winding condensate CFT.

\subsection{Convergence and analytic continuation}

In this subsection, we analyze the convergence properties of integrals of the form (\ref{CorFunGen}). First, we discuss nested divergences of Selberg-type integrals that arise naturally in Liouville theory. Then we will explore the mathematical connection between Selberg integrals and hypergeometric integrals and recover the results of nested divergences mentioned above from the point of view of the hypergeometric integral. In the next sub-section, we will apply these results to the specific context of two and three-point functions of the two winding CFT.

\subsubsection{Selberg integral}

For simplicity, we first look at the following simple integration (this integral and the more general ones in (\ref{LT3ptres}) arise as the residue of the three-point function in Liouville theory)
\begin{equation}
	\begin{aligned}
		G_1(a,b,0)=\int d^2 z |z|^{2a}|1-z|^{2b}= \int_{-\infty}^{+\infty} dx \int_{-\infty}^{+\infty} dy (x^2+y^2)^a((x-1)^2+y^2)^b 
	\end{aligned}
\end{equation}
Here we used $z=x+i y$, $x,y \in \mathbb{R}$. Now we analytically continue the integration in $y$ to a contour near the imaginary $y$ axis as follows
\begin{equation}\label{VarChange1}
	y \to iye^{-2i \epsilon}, ~~ \epsilon \to 0+
\end{equation}
Here $\epsilon$ serves as a regulator and tells us how the integration has to be performed near the singularities. We return to the variables  $z_\pm=x\pm y$ (analogous to holomorphic, anti-holomorphic) to write 
\begin{equation}
	\begin{aligned}
		G_1(a,b,0)=-\frac{i}{2}& \int_{-\infty}^{+\infty} dz_+  (z_+-i \epsilon (z_+-z_-))^a(z_+-1-i \epsilon (z_+-z_-))^b \\
		& \int_{-\infty}^{+\infty} dz_- (z_-+i \epsilon (z_+-z_-))^a(z_--1+i \epsilon (z_+-z_-))^b
	\end{aligned}
\end{equation}
For $z_+ \in (-\infty,0)$, the contour of integration over $z_-$ runs below the singularity at $z_-=0,1$. The integration contour can be deformed away near infinity on the lower half-plane and the contribution is zero provided we have 
\begin{equation}\label{UVCon}
	\Re[a+b]<-1
\end{equation}
A similar argument holds for $z_+ \in (1, \infty)$ (in which case the contour can be deformed to the upper half plane) and its contribution vanishes provided the condition above holds true. For $z_+ \in (0,1)$, the contour of integration over $z_-$ runs above the singularity at $z_-=0$ and below at $z_-=1$. We can deform the contour to run around $z_- \in (1, \infty)$ encircling the point at $z_- =1$. The integration over the circle around $z_- =1$ does not contribute anything provided 
\begin{equation}\label{IRCon}
	\Re[b]>-1
\end{equation}
In this case, the final answer comes from the integration over two lines $z_- \in (1, \infty)$   and takes the following form
\begin{equation}
	G_1(a,b,0)=-\sin (\pi b)\int_{0}^{1} dz_+  (z_+)^a (1-z_+)^b
 \int_{1}^{\infty} dz_-  (z_-)^a (z_--1)^b
\end{equation}
Integration over $z_+$ converges near $z_+=0,1$ provided
\begin{equation}
	\Re[a]>-1,~~ \Re[b]>-1 
\end{equation} 
We conclude that the integral representation of  $G(a,b)$ converges provided
\begin{equation}
	\Re[a]>-1,~~ \Re[b]>-1, ~~   \Re[a+b]<-1
\end{equation}
In fact, the result can easily be analytically continued away from this domain using a suitable definition of the single variable integrations above, for example 
\begin{equation}
	\int_{0}^{1} dz_+  (z_+)^a (1-z_+)^b=\frac{\Gamma(1+a)\Gamma(1+b)}{\Gamma(2+a+b)}
\end{equation} 
This gives\footnote{For the $z_-$ integral we change variables to $z_- \to 1/z_-$ and use the identity $\Gamma(z)\Gamma(1-z)=\pi/\sin(\pi z)$.}
\begin{equation}\label{G1ana}
	G_1(a,b,0)=\frac{\sin (\pi a)\sin (\pi b)}{\sin (\pi (a+b))}\(\frac{\Gamma(1+a)\Gamma(1+b)}{\Gamma(2+a+b)}\)^2
\end{equation}
From the answer, we can see that $\Gamma(1+a)$ remains finite as long as $1+\Re(a)>0$. Similarly, for  $\Gamma(1+b)$, we need  $1+\Re(b)>0$. These two also imply $\Re(a+b)>-2$. Note that the factor $\sin(\pi (a+b))$ will not hit a zero when
$-1>\Re(a+b)>-2$. Thus we recover all the conditions we discussed above as a consistency check.

We pause for a moment to understand something special about the analytic formula (\ref{G1ana}). Note that this explicit formula suggests that  as $b \to 0$ we have
\begin{equation}
	\lim_{b \to 0}G_1(a,b,0)= \mathcal{O}(b)
\end{equation}
 However, from the explicit integral representation, it is not obvious
\begin{equation}
	G_1(a,b,0)=- b \int_{0}^{1} dz_+  (z_+)^a 
 \int_{1}^{\infty} dz_-  (z_-)^a 
\end{equation}
because the integral has either UV or IR divergence. The explicit formula  (\ref{G1ana}) regularizes this divergence in a specific way. This fact will return in a very important way later in computing the two-point function.

Next we turn to the following  multiple integrals
\begin{equation}\label{LT3ptres}
	G_n(a,b,c)=\int \prod_{i=1}^n \[ d^2 z_i |z_i|^{2a}|1-z_i|^{2b}\] ~~\prod_{i<j} |z_i-z_j|^{2c} 
\end{equation}
We must be careful with nested divergences along with the divergences  discussed above. We can drop the contribution from large $z_{i,-}$ (keeping others fixed) provided
\begin{equation}
	\Re[a+b+(n-1)c]<-1
\end{equation}
We can ignore contributions from local singularities in the $z_{i,+}$ integration (with others staying away from any local singularity)  provided we demand
\begin{equation}
	\Re[a]>-1, \Re[b]>-1 \implies \Re(a+b)>-2
\end{equation}

Now we turn to nested divergences. For small $R$ we consider the situation when $z_{1,+}\approx z_{2,+}\approx \dots \approx z_{k,+} \approx R$, while we assume that the other $z_{i,+}$ are away from those points. In this case, the critical integration is from a place of the form
\begin{equation}
	\int dR ~ R^{k-1}R^{ka}R^{\frac{k}{2}(k-1)c}
\end{equation}
Convergence near the origin implies
\begin{equation}
	\Re(1+a+\frac{k-1}{2}c)>0, ~~ k=1,2,\dots, n \implies \Re( 1+a+\frac{n-1}{2}c)>0
\end{equation}
Similar consideration for  $z_{1,+}\approx z_{2,+}\approx \dots \approx z_{k,+} \approx 1-R$,  while assuming none of other $z_{i,+}$ are near any of those points, implies 
\begin{equation}
	\Re(1+b+\frac{k-1}{2}c)>0, ~~ k=1,2,\dots, n \implies \Re( 1+b+\frac{n-1}{2}c)>0
\end{equation}
Now consider the case when $z_{1,+}\approx z_{2,+}\approx \dots \approx z_{k,+} \approx x-R$, and $x\gg 1$ while assuming all of the other $z_{i,+}$ are away from those points. Then, there might be two different kinds of divergence - one from the large value of the center of mass of the points and the other from the relative position of the points. We consider with them separately. First, we focus on the relative position. The critical integration is from a place of the form 
\begin{equation}
	\int dR ~ R^{k-2} R^{\frac{k}{2}(k-1)c}
\end{equation}
Convergence near the origin implies
\begin{equation}
	\Re(1+\frac{k}{2}c)>0, ~~ k=1,2,\dots, n \implies \Re( 1+\frac{n}{2}c)>0
\end{equation}
Now we deal with the motion of the center of mass and  consider the situation when $z_{1,+}\approx z_{2,+}\approx \dots \approx z_{k,+} \approx 1/R$,  whereas we assume none of other $z_{i,+}$ are near any of those points.  It is then convenient to change variables to $z'_{i,+}=1/z_{i,+}, i=1,2,..,k$, so that the relevant integral in this case takes the following form
\begin{equation}
	\int dR ~R^{k-1} \frac{1}{R^{2k}}  R^{-ka-kb} R^{-(\frac{k}{2}(k-1))c} R^{-k(n-k)c}
\end{equation}
Convergence near the origin requires
\begin{equation}
	\Re(-k-k(a+b)-(\frac{k}{2}(k-1))c-k(n-k)c)>0 \implies \Re( 1+a+b+\frac{k-1}{2}c+(n-k)c)<0
\end{equation}
Setting $k=n$ we get
\begin{equation}
	\Re(1+a+b+\frac{n-1}{2}c)<0
\end{equation}
Setting $k=1$ we get
\begin{equation}
	\Re(1+a+b+(n-1)c)<0
\end{equation}

When these conditions are met this integration can be evaluated as
\begin{equation}
	\begin{aligned}
		G_n(a,b,c)= & S_n(a,b,c)^2\frac{\prod_{i=1}^n \sin(\pi (a+(i-1)\frac{c}{2})) \sin(\pi (b+(i-1)\frac{c}{2})) \sin(\pi i\frac{c}{2}))}{n! \prod_{i=1}^n \sin(\pi (a+b+(n+i-2)\frac{c}{2}))\sin(\pi \frac{c}{2}))}\\
		=&  S_n(a,b,c) S_n(-2-a-b-(n-1)c,b,c) \prod_{i=1}^n \frac{ \sin(\pi (b+(i-1)\frac{c}{2})) \sin(\pi i\frac{c}{2})}{\sin(\pi \frac{c}{2})}
	\end{aligned}
\end{equation}
Here the Selberg integral is given by
\begin{equation}
	\begin{aligned}
		S_n(a,b,c)=&\int_{[0,1]^n} \prod_{i=1}^n \[ dx_i x_i^{a}(1-x_i)^{b}\] ~~\prod_{i<j} |x_i-x_j|^{c}\\
		=& \frac{\prod_{i=1}^n \Gamma(1+a+(i-1)\frac{c}{2}) \Gamma(1+b+(i-1)\frac{c}{2}) \Gamma(1+i\frac{c}{2})}{\prod_{i=1}^n \Gamma(2+a+b+(n+i-2)\frac{c}{2})\Gamma(1+\frac{c}{2})}
	\end{aligned}
\end{equation}
We discuss the convergence of each of the factors one by one.

The following factor never diverges (essentially this factor takes care of various phases during contour manipulations as we will discuss momentarily)
\begin{equation}
 \prod_{i=1}^n \frac{ \sin(\pi (b+(i-1)\frac{c}{2})) \sin(\pi i\frac{c}{2})}{\sin(\pi \frac{c}{2})}
\end{equation}
Due to the identity 
\begin{equation}
	\begin{aligned}
	\frac{\sin(i \theta)}{\sin(\theta)}=2^{i-1}\prod_{k=1}^{i-1}\sin \( \theta+\frac{k\theta}{i}\)
	\end{aligned}
\end{equation}

In the expression of 
$S_n(a,b,c)$
the factor $\Gamma(1+a+(i-1)\frac{c}{2}) \Gamma(1+b+(i-1)\frac{c}{2})$ remains finite as long as we choose
\begin{equation}\label{S1}
\begin{aligned}
		& 1+\Re(a) >0, ~~ 1+\Re(b)>0\\
		& \Re(1+a+(n-1)\frac{c}{2})>0, ~~ \Re(1+b+(n-1)\frac{c}{2})>0 
		\end{aligned}
\end{equation}
The other factor to note for the purpose of diagnosing divergences is
\begin{equation}
	\frac{\Gamma(1+i\frac{c}{2})}{\Gamma(1+\frac{c}{2})}
\end{equation}
It is finite as long as 
\begin{equation}\label{S2}
	\Re\(\frac{c}{2}\)> -\frac{1}{n}
\end{equation}
Together (\ref{S1}) and (\ref{S2}) give a sufficient condition for the convergence of the Selberg integral $S_n(a,b,c)$ \cite{andrews_askey_roy_1999}.

The convergence of $S_n(-2-a-b-(n-1)c,b,c) $ in addition similarly requires
\begin{equation}\label{CS}
\begin{aligned}
		&  \Re(1+a+b+(n-1)\frac{c}{2})<0\\
		& \Re(1+a+b+(n-1)c)<0  
\end{aligned}
\end{equation}

Therefore we find complete agreement with the convergence analysis based on nested divergences. 

The fact that $G_n(a,b,c)$ is given by $ S_n(a,b,c) S_n(-2-a-b-(n-1)c,b,c) $ up to phase factors is easily understood as follows. As explained previously the integration gets non-zero contributions only when for all $i=1,2,.., n$  we have $0<z_{i+}<1$. For a given ordering of $z_{i,+}$s (say $0<z_{1,+}<z_{2,+}<\dots <z_{n,+}<1$) we have a particular contour for each $z_{i,-}$ running from $-\infty$ to $+\infty$ near the real line crossing the axis once in the interval $(0,1)$ (with the same ordering $0<z_{1,-}<z_{2,-}<\dots <z_{n,-}<1$ at the crossing point in $(0,1)$). The contour for all the  $z_{i,-}$ can be deformed without crossing each other to a contour that comes from $+\infty$ to $1$ below the real axis, circles at 1, and goes back to $+\infty$ above the real axis. Taking the branch cuts into account, the contribution from such contour can be mapped, up to phases, to one from $(0,1)$ by the inversion of all of the $z_{i,-}=1/z'_{i,-}$. Generically two halves of the contour would map \cite{Dotsenko:1984ad} to the same and opposite ordering of $z_{i,+}$ (in the case we are discussing  - it would be  $0<z_{1,-}<z_{2,-}<\dots <z_{n,-}<1$, $1>z_{1,-}>z_{2,-}>\dots >z_{n,-}>0$). The inversion effectively changes $(a,b,c)$ to $(-2-a-b-(n-1)c,b,c)$ for the $z_{i,-}$ integration. As a result $G_n(a,b,c)$ is the sum (with suitable phases) of the product of two Selberg integrals.

\subsubsection{Hypergeometric integral}

In next sub-section we will express Selberg type integrals in terms of  hypergeometric integrals \cite{GKZ} of the following form
\begin{equation}
	\begin{aligned}
	I_n(A,B,C)
	=& \int_{(0,1)^n} \prod_{i=1}^n \[ dx_i (x_i)^{A_i} \ (1-x_{i})^{B_i}\] ~~\prod_{1\le i<j \le n}  (1-X_{i,j}) ^{ C_{ij} }, ~~  X_{i,j}=x_i x_{i+1} \dots x_{j}
	\end{aligned}
\end{equation}
The convergence and analyticity of the integrals can be most conveniently understood in terms of the following variables
\begin{equation}\label{varC2}
	\begin{aligned}
	& A_i= s_{2,i+3}, ~~~  B_i=s_{i+2,i+4} \\
	& C_{ij}=s_{i+2,i+5}-s_{i+2,i+4}-s_{i+3,i+5}-1 ~~ \text{for }j=i+1\\
	& C_{ij}=s_{i+3,j+3}+s_{i+2,j+4}-s_{i+2,j+3}-s_{i+3,j+4} ~~ \text{for }j>i+1\\
	\end{aligned}
\end{equation}
The integral $I_n$ is an analytic function of $s_{ij}$ in the domain (see corollary 8.5 in \cite{brown2006multiple})
\begin{equation}
	\begin{aligned}
	1+\Re(s_{i,j}(a,b,c))>0
	\end{aligned}
\end{equation}
Here we need to impose the condition for all pairs $(i,j)$ that are defined in (\ref{varC2}). The Taylor expansion coefficients can be written in terms of multiple zeta values at non-negative integer values of $s_{i,j}$.

\subsubsection{Selberg to hypergeometric}

Consider the following integral
\begin{equation}
	\begin{aligned}
	I_n^{+}(a,b,c)=	\int_{0<u_1<u_2<\dots <u_n<1} \prod_{i=1}^n \[ du_i \ u_i^{a_i}(1-u_i)^{b_i}\] ~~\prod_{i<j} (u_j-u_i)^{c_{ij}} 
	\end{aligned}
\end{equation}
We can convert it to an integral of hypergeometric type by changing the variables to cubical ones ($x$)\footnote{Note that the ordering of $u_i$ fixes this substitution. For instance $-$ ordering ${1>u_1>u_2>\dots >u_n>0}$ corresponds to the  substitution 
$
		u_i=X_{1,i}
	$
}
\begin{equation}
	\begin{aligned}
		u_i=X_{i,n}
	\end{aligned}
\end{equation}
The integral above takes the following form
\begin{equation}
	\begin{aligned}
	I_n^{+}(a,b,c)=&	\int_{(0,1)^n} \prod_{i=1}^n \[ dx_i (x_i)^{i-1} \ X_{i,n}^{a_i}(1-X_{i,n})^{b_i}\] ~~\prod_{1\le i<j \le n}X_{j,n}^{c_{ij}} (1-X_{i,j-1})^{c_{ij}} \\
	=  & \int_{(0,1)^n} \prod_{i=1}^n \[ dx_i (x_i)^{A_i} \ (1-x_{i})^{B_i}\] ~~\prod_{1\le i<j \le n}  (1-X_{i,j}) ^{ C_{ij} } 
	\end{aligned}
\end{equation}
Here we defined
\begin{equation}\label{varC1}
	\begin{aligned}
	& A_i=i-1+\sum_{k=1}^{i} a_k+ \sum_{k=2}^{i}\sum_{l=1}^{k-1}c_{lk} ~~ \text{ for }1 \le i\le n \\
	& B_i= c_{i \ i+1} \text{ for } 1\le i \le n-1, ~~ b_n \text{ for } i=n \\
	& C_{ij}=c_{i ~ j+1} \text{ for }1\le  i<j\le n-1, ~  b_i \text{ for } 1\le i<j=n
	\end{aligned}
\end{equation}
This  determines $s^+_{ij}$ through (\ref{varC1}), (\ref{varC2}).

Note that the ordering is important. Therefore sufficient conditions for the Selberg integral  
\begin{equation}
	\begin{aligned}
		S_n(a,b,c)=&\int_{[0,1]^n} \prod_{i=1}^n \[ du_i u_i^{a}(1-u_i)^{b}\] ~~\prod_{i<j} |u_i-u_j|^{c}
	\end{aligned}
\end{equation}
to converge would be to impose 
\begin{equation}\label{HS1}
	\begin{aligned}
	1+\Re(s^\sigma_{i,j}(\{a_i\},\{b_i\},\{c_{ij}\}))>0
	\end{aligned}
\end{equation}
for all possible orderings $\sigma$ of $\{u_i\}$. However in practice due to the large amount of symmetry of the integrand, one might hope to get a sufficient condition by just imposing the constraint for $+,-$ orderings
\begin{equation}
	\begin{aligned}
	1+\Re(s^+_{i,j}(\{a_i\},\{b_i\},\{c_{ij}\}))>0, ~~ 1+\Re(s^+_{i,j}(\{b_i\},\{a_i\},\{c_{ij}\}))>0
	\end{aligned}
\end{equation}
They are related by $u_i \to 1-u_i$ change of variables that is a symmetry of $S_n(a,b,c)$. It turns out that these two conditions are equivalent and sufficient, i.e., these are the same as   (\ref{S1}) and (\ref{S2}).

Generalizing the arguments presented before, for the convergence of 
\begin{equation}
	G_n(a,b,c)=\int \prod_{i=1}^n \[ d^2 z_i |z_i|^{2a_i}|1-z_i|^{2b_i}\] ~~\prod_{i<j} |z_i-z_j|^{2c_{ij}} 
\end{equation}
we demand
\begin{equation}
	\begin{aligned}
	1+\Re(s^\sigma_{i,j}(\{a_i\},\{b_i\},\{c_{ij}\}))>0, ~~ & 1+\Re(s^{\pm\sigma}_{i,j}(\{-2-a_i-b_i-\sum_{n\ge j>i}c_{ij} \},\{b_i\},\{c_{ij}\}))>0
	\end{aligned}
\end{equation}
for all possible orderings $\sigma$. The domain of convergence constructed this way has the symmetries of the integral only after summing over all orderings $\sigma$. In practice for integrals of the type (\ref{CorFunGen}) we will impose (they are related by change of variables $z_i \to 1-z_i$)
\begin{equation}\label{doc1}
	\begin{aligned}
	1+\Re(s^+_{i,j}(\{a_i\},\{b_i\},\{c_{ij}\}))>0, ~~ 1+\Re(s^+_{i,j}(\{b_i\},\{a_i\},\{c_{ij}\}))>0
	\end{aligned}
\end{equation}
and (related to the above ones by combination of $z_i \to 1/z_i$ and $z_i \to 1-z_i$)
\begin{equation}\label{doc2}
	\begin{aligned}
	& 1+\Re(s^+_{i,j}(\{-2-a_i-b_i-\sum_{n\ge j>i}c_{ij} \},\{b_i\},\{c_{ij}\}))>0\\
	& 1+\Re(s^+_{i,j}(\{-2-a_i-b_i-\sum_{n\ge j>i}c_{ij} \},\{a_i\},\{c_{ij}\}))>0\\
	& 1+\Re(s^+_{i,j}(\{a_i\}, \{-2-a_i-b_i-\sum_{n\ge j>i}c_{ij} \},\{c_{ij}\}))>0\\
	& 1+\Re(s^+_{i,j}(\{b_i\}, \{-2-a_i-b_i-\sum_{n\ge j>i}c_{ij} \}, \{c_{ij}\}))>0
	\end{aligned}
\end{equation}

\newpage

\subsection{Two point function}\label{2point}
In this subsection, we consider the two-point function of operators that carry zero winding and momentum in both 
the circles. First, we will show by a formal change of variables that as long as the integral representation in (\ref{CorFunGen}) makes sense we have a power law dependence on the distance between the operators.  Then we will show the two-point functions of  non-identical operators vanish, which implies that one-point functions of non-trivial operators are zero. These results are non-trivial checks that the double condensate produces a genuine conformal field theory. 

We consider identical operators for the time being and set $w_1=0,w_2=R$ in (\ref{CorFunGen}).  We proceed to present a formal argument that the dependence on $R$ is indeed a power law. As a first step we change integration variables to $z_i, z_{i'}, z_I, z_{I'} \to Rz_i, Rz_{i'}, Rz_I, Rz_{I'}$. 
 
 (i) Factors of $R$ from the first line and the factor involving $w_2$ on 3rd, 4th, 8th, 9th give
 \begin{equation}
 	R^{-4\alpha^2+2j(2s_1+2s_2)}= R^{-2(2j^2 Q^2-2j(1+2j)Q^2) }=R^{-2(-2Q^2j(j+1)) }
 \end{equation} 
This gives the conformal dimension of external operators. Now we will show that the rest of the terms does not give rise to any new power of $R$.

(ii) Factors of $R$ from the second and 7th line and the factor involving $w_1$ on 3rd, 4th, 8th, 9th give
\begin{equation}
	R^{2s_1(2+2j)+2s_2(2+2j)}=R^{2(s_1+s_2)(1+(s_1+s_2)(k-2))}
\end{equation}

(iii) Factors of $R$ from  5th, 6th, 10th, 11th give
\begin{equation}
	R^{4s_1(s_1-1)/2-2(k-1)s_1^2+4s_2(s_2-1)/2-2(k-1)s_2^2}=R^{-2(s_1+s_2)-2(k-2)(s_1^2+s_2^2)}
\end{equation}

(iv) Factors of $R$ from  last four line give
\begin{equation}
	R^{8s_1s_2(-2b^2)}=R^{-4 s_1 s_2 (k-2)}
\end{equation}

Note that factors of $R$ from (ii)-(iv) cancel, proving the claim. This allows us to take $R \to \infty $ and evaluate the residue in a simple manner.

We expect the two-point function to be proportional to a delta function. At the level of residues, this implies the residue should diverge. We will carefully analyze this point below and show that this is indeed the case. We set $w_1=0$, and $w_2=R\to \infty$ above and strip off the divergent power of $R$ (which determines the worldsheet conformal dimension of the operator) to give
\begin{equation}\label{2ptfunres}
	\begin{aligned}
		&~~\res_{s\to 2s_1+2s_2} \langle  \tilde{V}^{0}_{\alpha,0,0}(0,0) \tilde{V}^{0}_{\alpha',0,0}(R,R)  \rangle\\
		=& ~~\frac{\pi}{b} \frac{(-1)^s}{s!} \sum_{2s_1+2s_2=s} \frac{(2s_1+2s_2)!}{(s_1!)^2 (s_2!)^2} \frac{\mu^{2s_1}}{\alpha_+^{4s_1}} \frac{\lambda^{2s_2}}{\alpha_+^{4s_2}} \  (N^{0,0}_{\alpha,0,0}N^{0,0}_{\alpha',0,0}) 
		~~ \\&~~~~ \int\prod_{i=1}^{s_1}  d^2z_i \int \prod_{i'=1}^{s_1}  d^2z'_{i'} \\
		&~~~~\prod_{i=1}^{s_1} \[ |z_i|^{2j}  \]
		~~~~\prod_{i'=1}^{s_1}  \[ |z'_{i'}|^{2j}  \]	~~\prod_{i'<j'}  \[ |z_{i'j'}|^2  \] ~~\prod_{i<j}   \[ |z_{ij}|^2 \]\ ~~  \prod_{i,j'}   \[ |z_{ij'}|^{2-2k}  \]\\
		&~~~~\\
		&~~~~\int\prod_{I=1}^{s_2}  d^2z_I \int \prod_{I'=1}^{s_2}  d^2z'_{I'} \\
		&~~~~\prod_{I=1}^{s_2} \[ |z_I|^{2j}  \]
		~~~~\prod_{I'=1}^{s_2}  \[ |z'_{I'}|^{2j}  \]	~~\prod_{I'<J'}  \[ |z_{I'J'}|^2  \] ~~\prod_{I<J}   \[ |z_{IJ}|^2 \]\ ~~  \prod_{I,J'}   \[ |z_{IJ'}|^{2-2k}  \]\\
		&~~~\\
		&~~~~ \prod_{i,J} \[ (z_{iJ})^{-\frac{k-2}{2}}  (\bar{z}_{iJ})^{-\frac{k-2}{2}}\]~~~~ \prod_{i',J} \[ (z_{i'J})^{-\frac{k-2}{2}}  (\bar{z}_{i'J})^{-\frac{k-2}{2}}\]\\
		&~~~~ \prod_{i,J'} \[ (z_{iJ'})^{-\frac{k-2}{2}}  (\bar{z}_{iJ'})^{-\frac{k-2}{2}}\]~~~~ \prod_{i',J'} \[ (z_{i'J'})^{-\frac{k-2}{2}}  (\bar{z}_{i'J'})^{-\frac{k-2}{2}}\]\\
	\end{aligned}
\end{equation}
The formula should be understood as follows. For a given value of $k, j$, the value of $j'$ is determined from knowledge of $s$.  Instead of analyzing all nested divergences, we will discuss specific cases only. 
 For small $R$ we consider the situation when $z_{1,+}\approx z_{2,+}\approx \dots \approx z_{l,+} \approx z'_{1,+}\approx z'_{2,+}\approx \dots \approx z'_{l',+} \approx R$. For convergence we need
 \begin{equation}\label{2ptAllSmall}
	(l+l')(1+j)+\frac{l}{2}(l-1)+\frac{l'}{2}(l'-1)+ll'(1-k)>0
\end{equation}
 For small $R$ we consider the situation when $z_{1,+}\approx z_{2,+}\approx \dots \approx z_{l,+} \approx z'_{1,+}\approx z'_{2,+}\approx \dots \approx z'_{l',+} \approx 1/R$.  Convergence requires 
 \begin{equation}\label{2ptAllLarge}
	-(l+l')(1+j)-(\frac{l}{2}(l-1)+\frac{l'}{2}(l'-1))+l^2+l'^2+(1-k)ll'+(k-2)(l+l')(s_1+s_2)>0
\end{equation}
For small $R$ we consider the situation when $z_{1,+}\approx z_{2,+}\approx \dots \approx z_{l,+} \approx z'_{1,+}\approx z'_{2,+}\approx \dots \approx z'_{l',+} \approx x+R$ with generic x.  Convergence of the integral in relative co-ordinate demands
 \begin{equation}\label{mid2}
	(l+l')+\frac{l}{2}(l-1)+\frac{l'}{2}(l'-1)+ll'(1-k)-1>0
\end{equation}
The simplest case is when $l=l'=1$. This gives from (\ref{mid2})
 \begin{equation}\label{conK1}
	2>k
\end{equation}
We add (\ref{2ptAllSmall}) with (\ref{2ptAllLarge}) and set $l=l'$ to obtain
 \begin{equation}\label{conK2}
(k-2)	2l(-l+s_1+s_2)>0 \implies k>2
\end{equation}

It is impossible to satisfy both (\ref{conK1}) and (\ref{conK2}). Therefore the integral (\ref{2ptfunres}) never converges. This fact can be separately verified using (\ref{doc1}) and (\ref{doc2}) with 
\begin{equation}
\begin{aligned}
& a_i=j, ~~ b_i=0, ~~ n=2s_1+2s_2\\
& c_{ij} = 1 ~~ \text{for  } (i,j) \in (1,s_1) \ \text{or} \ (s_1+1,2s_1)  \ \text{or} \  (2s_1+1,2s_1+s_2) \ \text{or} \ (2s_1+s_2+1,2s_1+2s_2) \\
&  c_{ij} = 1-k ~~ \text{for  } i \in (1,s_1) \text{ and } j \in  (s_1+1,2s_1)\\
&  c_{ij} = 1-k ~~ \text{for } \ i \in (2s_1+1,2s_1+s_2) \text{ and } j\in (2s_1+s_2+1,2s_1+2s_2) \\
&  c_{ij} = 1-\frac{k}{2} ~~ \text{for } \ i \in (1,2s_1) \text{ and } j\in (2s_1+1,2s_1+2s_2)
\end{aligned}
\end{equation}

However, naive contour manipulation would suggest we can always push the $z_{i,-}$ contour towards large values, and the integral would then vanish. Therefore we cannot conclude anything from this analysis alone. To understand the situation in more detail, we first discuss an analogous situation in Liouville theory (for our conventions see sub-section \ref{regInt}). The exact three-point function of $V_\alpha$s of Liouville theory is given by (\cite{Fateev:2007qn})
\begin{equation}
	\begin{aligned}
	C(\alpha_1,\alpha_2,\alpha_3)=\(\pi \mu \gamma(b^2)b^{2-2b^2} \)^{\frac{Q-\alpha}{b}} \frac{ Y'(0)}{Y(\alpha-Q)}\prod_{k=1}^3 \frac{Y(2\alpha_k)}{Y(\alpha-2\alpha_k)}, ~~ \alpha=\sum_{k=1}^3 \alpha_k
	\end{aligned}
\end{equation}
Where the function
 \begin{equation}
	\begin{aligned}
	Y(x)=\text{Exp}\[\int_0^\infty \frac{dt}{t}\(\(\frac{Q}{2}-x\)^2e^{-t}-\frac{
	\sinh^2(\frac{Q}{2}-x)\frac{t}{2}}{\sinh \frac{bt}{2} \sinh \frac{t}{2b}} \)\]
	\end{aligned}
\end{equation}
has first-order zeros at
\begin{equation}
	\begin{aligned}
		x=-m b - \frac{n}{b}, \ Q+m b +\frac{n}{b}~~~ \text{for }m,n=0,1,2,..
	\end{aligned}
\end{equation}
The two-point function obtained by $\alpha_2\to 0$ in Liouville theory has a delta function $\delta (\alpha_1-\alpha_3)$. We can show this by noting that 
\begin{equation}\label{propLT}
	\begin{aligned}
		& ~~~\lim_{\alpha_2\to 0} C(\alpha_1,\alpha_2,\alpha_3) = \mathcal{O}(\alpha_2)\\
		& \lim_{\alpha_3\to \alpha_1+\alpha_2} C(\alpha_1,\alpha_2,\alpha_3) =\mathcal{O}((-\alpha_3+\alpha_1+\alpha_2)^{-1})
	\end{aligned}
\end{equation} 
 This follows by keeping track of zeros of $Y$ appearing in the expression of the three-point function. As a result, we obtain
 \begin{equation}
 	\begin{aligned}
 	\lim_{\epsilon \to 0}	\lim_{\alpha_2\to \epsilon} C(\alpha_1,\alpha_2,\alpha_3) =\mathcal{O}\( \lim_{\epsilon \to 0} \frac{\epsilon}{(\alpha_1-\alpha_3+\epsilon) (\alpha_3-\alpha_1+\epsilon)} \)
 	\end{aligned}
 \end{equation}
When $\alpha_1-\alpha_3 = i (s_1-s_3)$ for real $s_1,s_3$ we get a delta function as claimed above.  Now we turn to see the key properties (\ref{propLT}) from the residues of the three-point function itself. The point of this exercise is to show that the relevant properties have their origin  in perturbative calculations.
 \begin{equation}
 	\begin{aligned}
 		\lim_{\alpha \to Q-n b} C(\alpha_1,\alpha_2,\alpha_3) = (-\pi \mu)^n R_n(\alpha_1,\alpha_2,\alpha_3)
 	\end{aligned}
 \end{equation}
Where
  \begin{equation}\label{resLT1}
 \begin{aligned}
R_n(\alpha_1,\alpha_2,\alpha_3)
&= \prod_{k=1}^n\(\frac{\gamma(-kb^2)}{\gamma(-b^2)}\)\ \prod_{j=0}^{n-1}\(\frac{1}{\gamma(2b\alpha_1+j b^2)\gamma(2b\alpha_2+j b^2)\gamma(2b\alpha_3+j b^2)}\)
 	\end{aligned}
 \end{equation}
We should think of (\ref{resLT1}) as a function of $\alpha_1, \alpha_2, n$. The variable $\alpha_3=Q-nb-\alpha_1-\alpha_2$ is completely determined in terms of them. We note from the explicit expression of $R_n$ that
\begin{equation}\label{propLTres}
	\begin{aligned}
		& ~~~\lim_{\alpha_2\to 0} R_n(\alpha_1,\alpha_2,\alpha_3) = \mathcal{O}(\alpha_2)\\
		& \lim_{\alpha_3\to \alpha_1+\alpha_2} R_n(\alpha_1,\alpha_2,\alpha_3) =\mathcal{O}((-\alpha_3+\alpha_1+\alpha_2)^{-1})
	\end{aligned}
\end{equation}  
Now that we know exactly what properties of the function $R_n$ we are looking for, we turn into its integral representation, which can be obtained from the path integral 
\begin{equation}\label{resLT2}
 \begin{aligned}
R_n(\alpha_1,\alpha_2,\alpha_3)&=\int \prod_{j=1}^{n}[d^2z_j] \prod_{i<j}\[ |z_i-z_j|^{-4b^2}\]\int \prod_{j=1}^{n}\[|z_j|^{-4b\alpha_1} |1-z_j|^{-4b\alpha_2}\] 	\end{aligned}
 \end{equation}
 Convergence of this integral demands\footnote{It's same as $G_n(-2b \alpha_1,-2b \alpha_2,-2b^2)$. The upper bound on $\alpha_2$ can be seen from (\ref{S}), (\ref{CS}).}
 \begin{equation}
 	\begin{aligned}
 		-1<-2b \alpha_1<0, ~~-1<-2b \alpha_2<0, ~~ 1-2b \alpha_1-2b \alpha_2-(n-1)b^2<0
 	\end{aligned}
 \end{equation}
As $\alpha_2 \to 0$ we naively face a divergence, but from the explicit formula above we see that this divergence must be regulated, as we should get zero. This is the same phenomenon as seen in the example of $G_1(a,b,0)$ previously. Now we turn to the second property in (\ref{propLTres}). For $\alpha_3= \alpha_1+\alpha_2 + \epsilon$ we have $\alpha_1+\alpha_2=(Q-nb-\epsilon)/2 $. As a result 
\begin{equation}
	1-2b \alpha_1-2b \alpha_2-(n-1)b^2=\epsilon b
\end{equation}
Thus we end up getting a divergence as expected. The divergence can be easily seen by looking at the integral representation and focusing on the region where $z_i \sim 1/R$ are large, so the structure of the integral \ref{resLT2} is 
\begin{equation}
	\int dR R^{-1+2n(1-b^2(n-1) -2b (\alpha_1+\alpha_2)) } = \int dR R^{-1+2n b \epsilon}= \mathcal{O}(\epsilon^{-1})
\end{equation} 
 This is in agreement with the second property in (\ref{propLTres}).

 With this understanding, we turn to (\ref{2ptfunres}). We expect that with suitable regularization, the residue in (\ref{2ptfunres}) for generic values of $j$ (unrelated to $s$) would vanish as expected from naive manipulation of the contours.  We turn to the discussion of the analog of the second property in (\ref{propLTres}) in the next section. 
 
 \label{fig1}
\begin{figure}[ht]
  \subfloat[]{
	\begin{minipage}[c][1\width]{
	   0.3\textwidth}
	   \centering
	   \includegraphics[width=1\textwidth]{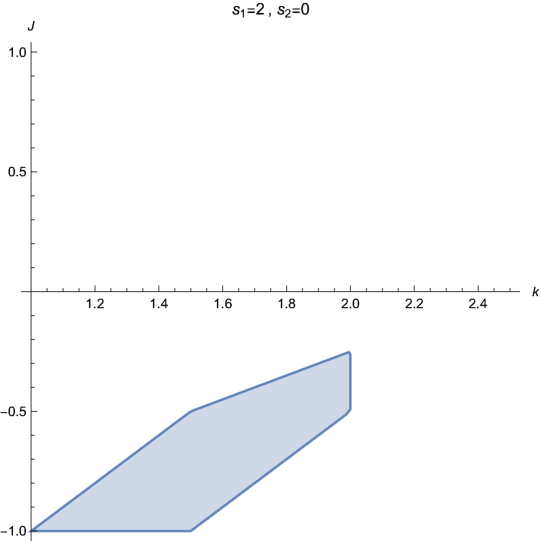}
	\end{minipage}}
 \hfill 	
  \subfloat[]{
	\begin{minipage}[c][1\width]{
	   0.3\textwidth}
	   \centering
	   \includegraphics[width=1.1\textwidth]{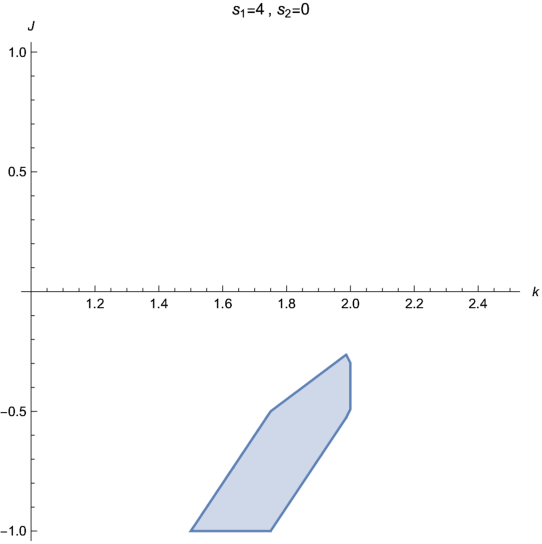}
	\end{minipage}}
 \hfill	
  \subfloat[]{
	\begin{minipage}[c][1\width]{
	   0.3\textwidth}
	   \centering
	   \includegraphics[width=1.2\textwidth]{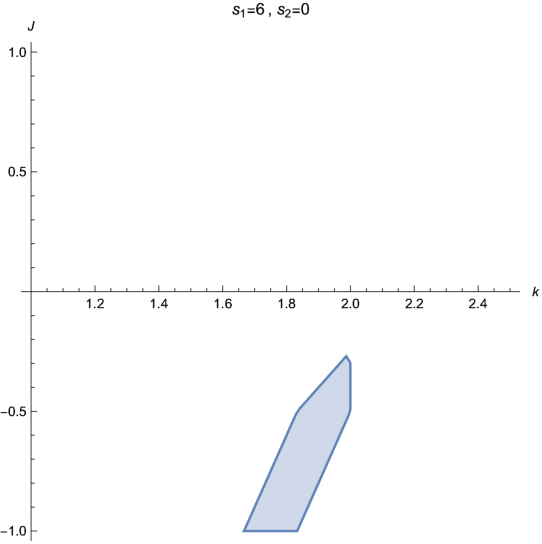}
	\end{minipage}}
\caption{Domain of convergence of the residue of the three-point function for $j_1=j_2=j$ on $j$ vs $k$ plane. Note that the domain is the same for $(s_1,s_2)=(s_2,s_1)$ which we do not depict.}
\end{figure}

 \begin{figure}[ht]
  \subfloat[]{
	\begin{minipage}[c][1\width]{
	   0.3\textwidth}
	   \centering
	   \includegraphics[width=1\textwidth]{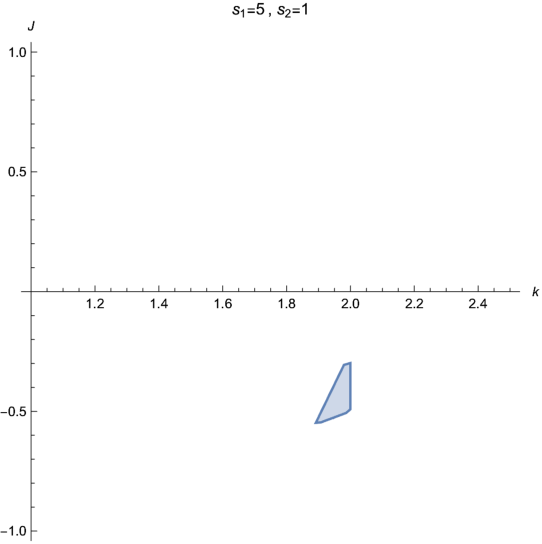}
	\end{minipage}}
 \hfill 	
  \subfloat[]{
	\begin{minipage}[c][1\width]{
	   0.3\textwidth}
	   \centering
	   \includegraphics[width=1.1\textwidth]{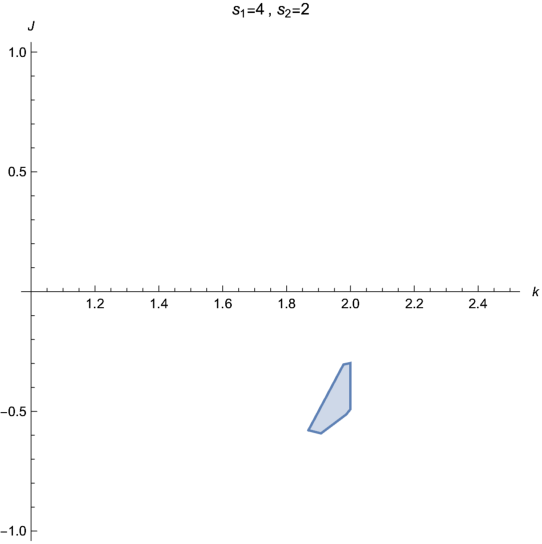}
	\end{minipage}}
 \hfill	
  \subfloat[]{
	\begin{minipage}[c][1\width]{
	   0.3\textwidth}
	   \centering
	   \includegraphics[width=1.2\textwidth]{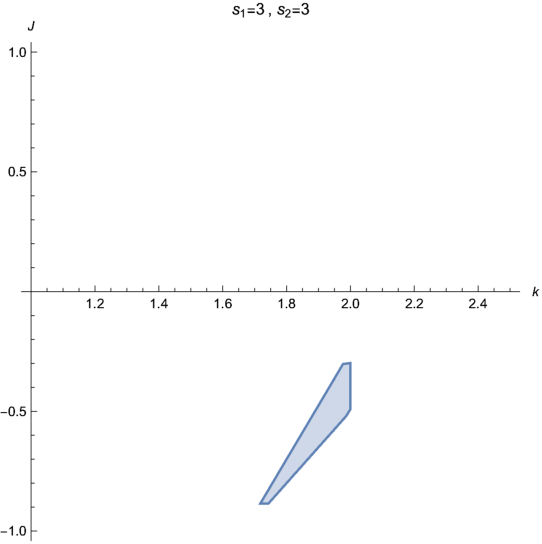}
	\end{minipage}}
\caption{Domain of convergence of the residue of the three-point function for $j_1=j_2=j$ on $j$ vs $k$ plane. }
\label{fig2}
\end{figure}

\subsection{Three point function}

In this sub-section, we focus on the integral representation for the residue of the three-point function for the Double winding condensate CFT. We will show that there is a non-empty domain of convergence in $k,j_1,j_2,j_3$ space for the integral representation in (\ref{CorFunGen}). Therefore we can analytically continue away from this domain to general values in a manner similar to the single winding condensate CFT.  Arguments similar to those we used for two point function show that whenever the integral representation for the residue of the three-point function is convergent, we formally have the structure of the three-point function of a CFT. In addition, we will examine the two properties in (\ref{propLTres}). For one winding CFT, the properties will come out from the analysis of the domain of convergence. We will discuss the subtleties involved when both windings are turned on.

We set $w_1=0, w_2=1$, and $w_3=R\to \infty$ above and strip off the divergent power of $R$ (which determines the worldsheet conformal dimension of the operator) to give
\begin{equation}
	\begin{aligned}
		&~~\res_{s\to 2s_1+2s_2} \langle   \tilde{V}^{0}_{\alpha_1,0,0}(0,0) \tilde{V}^{0}_{\alpha_2,0,0}(1,1) \tilde{V}^{0}_{\alpha_3,0,0}(R,R)  \rangle\\
		=& ~~\frac{\pi}{b} \frac{(-1)^s}{s!} \sum_{2s_1+2s_2=s} \frac{(2s_1+2s_2)!}{(s_1!)^2 (s_2!)^2} \frac{\mu^{2s_1}}{\alpha_+^{4s_1}} \frac{\lambda^{2s_2}}{\alpha_+^{4s_2}} \  (N^{0,0}_{\alpha_1,0,0}N^{0,0}_{\alpha_2,0,0}N^{0,0}_{\alpha_3,0,0}) 
		~~ \\&~~~~ \int\prod_{i=1}^{s_1}  d^2z_i \int \prod_{i'=1}^{s_1}  d^2z'_{i'} \\
		&~~~~\prod_{i=1}^{s_1} \[ |z_i|^{2j_1} |z_i-1|^{2j_2}  \]
		~~~~\prod_{i'=1}^{s_1}  \[ |z'_{i'}|^{2j_1} |z'_{i'}-1|^{2j_2}  \]	~~\prod_{i'<j'}  \[ |z_{i'j'}|^2  \] ~~\prod_{i<j}   \[ |z_{ij}|^2 \]\ ~~  \prod_{i,j'}   \[ |z_{ij'}|^{2-2k}  \]\\
		&~~~~\\
		&~~~~\int\prod_{I=1}^{s_2}  d^2z_I \int \prod_{I'=1}^{s_2}  d^2z'_{I'} \\
		&~~~~\prod_{I=1}^{s_2} \[ |z_I|^{2j_1} |z_I-1|^{2j_2} \]
		~~~~\prod_{I'=1}^{s_2}  \[ |z'_{I'}|^{2j_1} |z'_{I'}-1|^{2j_2}  \]	~~\prod_{I'<J'}  \[ |z_{I'J'}|^2  \] ~~\prod_{I<J}   \[ |z_{IJ}|^2 \]\ ~~  \prod_{I,J'}   \[ |z_{IJ'}|^{2-2k}  \]\\
		&~~~\\
		&~~~~ \prod_{i,J} \[ (z_{iJ})^{-\frac{k-2}{2}}  (\bar{z}_{iJ})^{-\frac{k-2}{2}}\]~~~~ \prod_{i',J} \[ (z_{i'J})^{-\frac{k-2}{2}}  (\bar{z}_{i'J})^{-\frac{k-2}{2}}\]\\
		&~~~~ \prod_{i,J'} \[ (z_{iJ'})^{-\frac{k-2}{2}}  (\bar{z}_{iJ'})^{-\frac{k-2}{2}}\]~~~~ \prod_{i',J'} \[ (z_{i'J'})^{-\frac{k-2}{2}}  (\bar{z}_{i'J'})^{-\frac{k-2}{2}}\]\\
	\end{aligned}
\end{equation}
Instead of giving a detailed analysis of nested divergences, we will rely on  (\ref{doc1}) and (\ref{doc2}) with 
\begin{equation}
\begin{aligned}
& a_i=j_1, ~~ b_i=j_2, ~~ n=2s_1+2s_2\\
& c_{ij} = 1 ~~ \text{for  } (i,j) \in (1,s_1) \ \text{or} \ (s_1+1,2s_1)  \ \text{or} \  (2s_1+1,2s_1+s_2) \ \text{or} \ (2s_1+s_2+1,2s_1+2s_2) \\
&  c_{ij} = 1-k ~~ \text{for  } i \in (1,s_1) \text{ and } j \in  (s_1+1,2s_1)\\
&  c_{ij} = 1-k ~~ \text{for } \ i \in (2s_1+1,2s_1+s_2) \text{ and } j\in (2s_1+s_2+1,2s_1+2s_2) \\
&  c_{ij} = 1-\frac{k}{2} ~~ \text{for } \ i \in (1,2s_1) \text{ and } j\in (2s_1+1,2s_1+2s_2)
\end{aligned}
\end{equation}
It can be checked that the resulting domain of convergence is symmetric under the exchange of $s_1,s_2$ as expected from the symmetry of the integral.

We first focus on the case of one winding condensate $s_2=0$. For many simple examples $s_1=1,2,...,6$ we present the  domain of convergence  below (here $i=1,2$)
\begin{equation}\label{3pt1winding}
	\begin{aligned}
		 2>k, ~ j_i>-1, ~~ s_1(k-2)>j_1+j_2, ~~j_i> \frac{(s_1(k-2)-1)}{2}>j_1+j_2
	\end{aligned}
\end{equation}
The corresponding domains are plotted in \ref{fig1}. We particularly focus on the last inequality
\begin{equation}
	\begin{aligned}
		 \frac{(s_1(k-2)-1)}{2}>j_1+j_2
	\end{aligned}
\end{equation}
Note that the condition to be on a pole whose residue is being computed can be written as
\begin{equation}\label{ineq2pt}
	\begin{aligned}
		(k-2)s_1=1+j_1+j_2+j_3
	\end{aligned}
\end{equation}
If we focus on the situation (as necessary to discuss the analog of  (\ref{propLTres}))
\begin{equation}
	\begin{aligned}
		j_3=j_1+j_2
	\end{aligned}
\end{equation}
The inequality in  (\ref{ineq2pt}) is precisely saturated, signaling a divergence. Therefore in this case, we expect the two-point function to have the structure of a delta function. In fact, this can be checked explicitly using the duality of the integrals.

Now we turn to the discussion of two winding condensates. We focus on the case of $s_1+s_2=6$ as an example. The domain of convergence for $s_1 \ge s_2 $ is given by
\begin{equation}
	\begin{aligned}
		 2>k, ~~ s_2(k-2)-1>2j_i, ~~  \frac{((2s_1+s_2)(k-2)-1)}{2}>j_1+j_2
	\end{aligned}
\end{equation}
The domain is plotted in fig. (\ref{fig2}). From the graphs it is clear that for a given set of external vertex operators, there exists a common domain of convergence for all the winding integrals for different values of $s_1,s_2$ (for given $s_1+s_2$) that contribute to the three point function. This demostrates that we can safely analytically continue from this domain to generic values to define the residue of the three point function.

Note that the condition to be on a pole whose residue is being computed can be written as
\begin{equation}
	\begin{aligned}
		(k-2)(s_1+s_2)=1+j_1+j_2+j_3
	\end{aligned}
\end{equation}
If we focus on the situation 
\begin{equation}
	\begin{aligned}
		j_3=j_1+j_2
	\end{aligned}
\end{equation}
We get 
\begin{equation}\label{3pt2winding}
	\begin{aligned}
		j_1+j_2= \frac{((s_1+s_2)(k-2)-1)}{2}>\frac{((2s_1+s_2)(k-2)-1)}{2}
	\end{aligned}
\end{equation}
Therefore when $s_1s_2 $ is non-zero, our integral representation breaks down before we can comment on the delta function necessary for the two-point function. 

\subsection{Regularity in the interior}\label{regInt}

We now turn to the additional conditions on the coupling constants $\mu, \lambda $ required for the two-winding condensate theory to be a sensible conformal field theory. These conditions originate from `regularity in the interior' in the target space language, given fixed AdS asymptotics.\footnote{In general, it is not clear whether there exists a well-defined CFT if vertex operators that are exponentially growing in the AdS asymptotic region are tuned on, corresponding to irrelevant deformations of a holographically dual theory on the boundary.} This implies that the reflection coefficients for the condensates themselves, as computed in the formalism we have outlined in the previous sections, must vanish. Since a sum over both types of winding operators appears in the calculations, this condition can only be satisfied for special values of $\mu/\lambda$.

In the winding condensate description of thermal AdS$_3$, one can use free field notation for the vertex operators, in terms of a $\beta \gamma$ operator times a linear dilaton $\varphi$ momentum operator, $e^{-2Qj \varphi }$, where $Q=1/\sqrt{k-2}$ is the background charge. 
In the free theory, there is another operator with the same quantum numbers, given by the reflection $j \rightarrow -1-j$. Passing to the interacting theory, these operators mix, in the sense that the physical vertex operator is a  reflection invariant combination of the two which diagonalizes the worldsheet CFT two-point function. 

From the point of view of the asymptotic region, one can interpret the two components as an incident and reflected wave. This is simply the reflection of the smooth origin of polar coordinates in the target space in the non-linear sigma model description, associated with regularity in the interior. For vertex operators associated with massive, non-tachyonic fields in Euclidean target space, there is always a reflected wave, and there are no normalizable such vertex operators. Thus, for example, at a generic temperature, the time winding marginal vertex operator in thermal AdS$_3$ contains a non-normalizable reflected wave. Here we are interested in translation invariant deformations (with respect to the two translation symmetries of Euclidean thermal AdS$_3$). 

In this paper, we investigated the theory obtained by adding both space and time winding operators to the free theory. It is clear that at general values of the ratio of the coefficients $\mu/\lambda $, the resulting theory cannot be a CFT on its own, without further inclusion of (at minimum) condensates of non-normalizable reflections of those operators, which would ruin the AdS asymptotics. Equivalently, the double winding condensate on its own would not be smooth in the interior. This corresponds to the lack of an interpolating conformal manifold at a fixed temperature.

We first discuss similar conditions in Liouville-like theories. A natural basis of operators in Liouville conformal field theory with central charge $ c=1+6 Q^2, Q=b+b^{-1},b>0$ is given by $O_\alpha$ satisfying the following equivalence\footnote{This is emphasized in \cite{Balthazar:2018qdv} in the context of $c=1$ string theory.}
\begin{equation}\label{normalizableOpL}
    O_{\alpha}=O_{Q-\alpha}, \quad \alpha=\frac{Q}{2}+iP, \quad P\in \mathbb{R}_{\geq0} %
\end{equation}
where the two-point function is canonically normalized 
\begin{equation}
    \langle O_{\alpha}(z,\bar{z}) O_{\alpha'}(0,0)\rangle= \frac{2\pi \delta(\alpha-\alpha')}{z^{2h_\alpha} \bar{z}^{2h_\alpha}}, \quad h_\alpha=\alpha(Q-\alpha)
\end{equation}
The theory can be completely defined as an abstract CFT in terms of the OPE of these operators \cite{Teschner:2001rv, Harlow:2011ny}. However, describing Liouville theory completely from the path integral  is a very subtle question as we explain below. 
The simplest attempt to this end is to define the theory in terms of the following action on the sphere \cite{Teschner:2001rv}
\begin{equation}\label{Liouville}
	\begin{aligned}
		& S=\frac{1}{4\pi}\int d^2\sigma \ \sqrt{\tilde{g}} \[ \tilde{g}^{\mu \nu} \partial_{\mu} \phi \bar{\partial}_{\nu}\phi +4 \pi\mu e^{2b \phi}+4 \pi \mu' e^{2b^{-1} \phi}+Q \tilde{R}\phi\]
	\end{aligned}
\end{equation}
and identify
\begin{equation}\label{Lop}
    O_{\alpha}=\frac{1}{2}\(R_L(\alpha)^{-\frac{1}{2}}V_{\alpha}+R_L(Q-\alpha)^{-\frac{1}{2}}V_{Q-\alpha}\), \quad V_{\alpha}=e^{2\alpha \phi}
\end{equation}
Here $R(\alpha)$ is the `reflection' coefficient satisfying
\begin{equation}
	R_L(Q-\alpha)=R_L(\alpha)^{-1}=\frac{b^2 (\pi \mu \gamma(b^2))^{(2\alpha-Q)/b}}{\gamma(2\alpha/b-1-1/b^2)\gamma(2\alpha b-b^2)}  
\end{equation}
By studying the quantum mechanics of the spatial zero modes of $\phi$ in the small $b \to 0$ limit, one can establish that for non-singular scattering in the interior ($\phi \to +\infty$) the wave function in the asymptotic $\phi \to -\infty$ region does indeed take the form in (\ref{Lop}) \cite{Seiberg:1990eb}. Therefore in the full theory, only the basis of operators in (\ref{Lop}) is well-defined, corresponding to the `criterion of regularity in the interior'.

The theory as given in (\ref{Liouville}) depends on three parameters $\mu, \mu',b$.\footnote{Under the field redefinition of $\phi$ by a constant shift, $\mu$ and $\mu'$ transform, together with the dilaton.} The strong-weak Liouville self-duality  fixes the coefficient $\mu'$  in terms of $\mu, b$, i.e., $\mu'=\mu'(\mu,b)$ as follows. The Lagrangian is invariant under the duality map $(\mu,\mu'(\mu,b),b)\to (\hat{\mu},\mu'(\hat{\mu},b^{-1}),b^{-1})$ provided we demand
\begin{equation}
    \begin{aligned}
        \hat{\mu}=\mu'(\mu,b), \quad \mu=\mu'(\hat{\mu},b^{-1})
    \end{aligned}
\end{equation}
This is solved by
\begin{equation}\label{Lduality}
    \mu'(\mu,b)=\frac{(\pi \mu \gamma(b^2))^{\frac{1}{b^2}}}{\pi \gamma(\frac{1}{b^2})}
\end{equation}
Therefore, Liouville theory can be thought of as a deformation of the linear dilaton theory by the duality invariant operator 
\begin{equation}
    O=\mu V_b+\mu'(\mu,b) V_{Q-b}\implies \frac{\partial}{\partial \mu}O=2R_L(b)^{\frac{1}{2}}O_b
\end{equation}
This operator does not belong to the set of normalizable operators $O_\alpha$ introduced previously in (\ref{normalizableOpL}).\footnote{Its two-point function is given by \begin{equation}
    \langle O(z,\bar{z}) O(0,0)\rangle= \frac{2\pi \delta(0)}{z^{2} \bar{z}^{2}}\mu \mu' Q^2
\end{equation}
}

The theory has an asymptotically free region near $\phi \to -\infty$, where the marginal operator $O$ decays exponentially. When the external operators obey $$Q-\sum_i\alpha_i<0,$$ the path integral over the $\phi$ zero mode converges. There is a pole when $Q -\sum \alpha_i=0$  from the asymptotic free region, whose residue is given by free field Wick contractions. Analytically continuing to the regime $Q-\sum \alpha_i>0$, the path integral is dominated by the free field region. Furthermore, when other resonance conditions for the external momenta are satisfied, for instance, $$Q-\sum_i\alpha_i=n b, \quad n\in \mathbb{Z}_{\geq0}$$ a particular Taylor expansion coefficient of the interactions (here from $O^n$) similarly leads to a free field calculation of the residue at the associated pole  \cite{Polyakov:1991qx}, \cite{Dorn:1994xn, Zamolodchikov:1995aa}. Therefore, in Liouville theory, both the incident and reflected wave of the Liouville operator are present in the theory, and both decay in the free field asymptotic region. 

The situation is similar for the cigar sigma model but with an important difference. There is a Liouville-like radial direction $\phi$ and a compact angular direction $\theta$, in terms of which the basis of operators that are regular in the interior takes the form (\ref{Lop}) near the boundary where the `free field' description is valid \cite{Giveon:1999tq, Aharony:2004xn, Jafferis:2020lgl}

\begin{equation}\label{Sop}
    \begin{aligned}
         & O_{j,m,n}=\frac{1}{2}\(R_C(j,m,n)^{-\frac{1}{2}} V_j+R_C(-1-j,m,n)^{-\frac{1}{2}} V_{-1-j} \), \quad V_{j}=e^{-2Qj \phi}\\
         & R_C(-1-j,m,n)=R_C(j,m,n)^{-1} \propto \frac{\Gamma(1+\frac{2j+1}{k-2})}{\Gamma(1-\frac{2j+1}{k-2})}  \frac{\Gamma(1+2j)}{\Gamma(-1-2j)}  \frac{\Gamma(-j+\frac{|m|-kn}{2})\Gamma(-j+\frac{|m|+kn}{2})}{\Gamma(1+j+\frac{|m|-kn}{2})\Gamma(1+j+\frac{|m|+kn}{2})}
    \end{aligned}
\end{equation}
Here $j$ is the radial momentum and $m,n$ are the momentum and winding in the compact circle. In the expression for the reflection coefficient, we have omitted a factor that is independent of the quantum numbers of the operator under consideration. The theory has an FZZ dual description in terms of a deformation of the free theory on the cylinder by a linear combination  of two operators $j=-k/2+1,m=0,n=\pm1$ \cite{Giveon:2016dxe}. Note that for sufficiently large $k$, these operators decay in the asymptotic weak coupling region $\phi\to -\infty$. 
To justify the free field calculation of the residues as in Liouville, all operators in the Lagrangian must decay in the free asymptotic region. 

The crucial difference  is that these winding operators now have reflected waves associated with $-j-1$ which do not decay in the $\phi\to -\infty$ limit.\footnote{This is true for $k\geq 4$. For $2<k<4$ both the winding condensate and the reflected wave of it decay in the asymptotic region $\phi \to -\infty$. However, to determine which one is normalizable in AdS one needs to take the change between the string frame and the Einstein frame into account. For normalizable fields in Euclidean signature 
\begin{equation}
    \lim_{\phi\to -\infty}  e^{-Q\phi} O_{j,m,n}=\text{finite}
\end{equation}
For $k>3$ the reflected wave from $j=-k/2+1,m=0,n=\pm1$ is not normalizable. For $k<3$, reflected waves are normalizable, but the original winding condensates are non-normalizable.  } The reflected waves for these operators are of the same form as in (\ref{Sop}) thanks to the FZZ duality. Amazingly, what happens instead is that the reflection coefficient  $R_C(-k/2+1,0,\pm1)$ simply vanishes, so there is no ``self-reflection'' of the condensate. This happens only when the radius of the compact boson and the linear dilaton background charge has a particular relation, which is precisely that which appears in the dual cigar background.

In the winding condensate description of the worldsheet in Euclidean AdS$_3$, the discussion is similar to the cigar sigma model above - the reflected waves from the winding condensates in Euclidean AdS$_3$ are absent because their reflection coefficient vanishes. For the double winding condensate theory presented in this paper, we do not know the exact formula for the reflection coefficient. It is important to note  here that the double winding condensate CFT at the Hawking-Page temperature is finitely away from the thermal AdS background. As a result, a discussion of reflected waves in the thermal AdS background is not relevant. On the other hand, as we increase the temperature to near Hagedorn we expect our theory to get corrections from other normalizable modes in thermal AdS to sensibly connect to a Horowitz-Polchinski-like solution. If there were reflected waves reaching all the way to the weak coupling region near the boundary of AdS, the free field calculations performed in this paper would be questionable.

We conjecture that \textit{the constraints from the `regularity in the interior' along with the Bootstrap constraints will fix the coupling constants $\mu, \lambda$ of the space- and time-winding condensates to produce a sensible conformal field theory}.\footnote{For related discussion see \cite{Kazakov:2000pm, Berkooz:2007fe}.} For $k>3$ the definitive test will be to compute the reflection amplitude from the two-point function as in section (\ref{2point}) for the condensate operators themselves, and determine where it vanishes as a function of $\mu/\lambda$ after summing the contributions from both types of winding operators. At $k=3$ the functional form of the reflected waves from the winding condensates is the same as that of the winding condensates itself. At this special value of $k$, one can  evaluate the string theoretic partition on toroidal worldsheet entirely based on the Lagrangian (\ref{doublewindingaction}) of this paper and demand consistent results based on holography to fix the ratio $\mu/\lambda$.\footnote{This is similar to the discussion of two dimensional string theory in \cite{PhysRevLett.65.3088}. We thank Igor Klebanov for pointing it to us. }   

\section{Future directions}\label{5}

In this paper, we have considered a two-dimensional theory built out of the free theory near the boundary of  AdS$_3$ deformed by winding condensates on both cycles of the boundary torus. 
We have analyzed the resulting theory \textit{exactly} at the Hawking-Page temperature and found substantial evidence that it is a conformal field theory. In particular, we have shown that one-point functions of non-trivial operators vanish and the two-point function is scale-invariant, and we presented an integral representation of the residue of all higher-point correlation functions. For the three-point function, we have analyzed the convergence of the integral representation in detail and found that Lorenzian worldsheet line integrals provide a better definition compared to Euclidean volume integrals. We have not studied the precise phase factors that appear in this transformation as a function of the choice of contours in the Lorentzian worldsheet. If the theory under consideration is an integrable one, one expects to find a natural organizing principle for these phases, as they appear in the four-point function for instance, by considering the Moore-Seiberg `classical' limit \cite{Moore:1988qv} (see also \cite{Alvarez-Gaume:1988bek}) of the conformal field theory. Further, one would expect to find null states dictated by a Knizhnik–Zamolodchikov type differential equation \cite{Knizhnik:1984nr}. Given the connection to hypergeometric integrals and the rich mathematical literature available for them  \cite{GKZ, brown2006multiple} (see \cite{GrobnerDeformation} for a pedagogical review), both these questions are intriguing directions for future investigation \cite{Barak}. 

 String perturbation theory around the BTZ black hole below a certain temperature develops a tachyonic instability involving a spatial winding operator. 
Switching the roles of space and time, string theory in thermal AdS$_3$ has an instability for a time winding tachyon above the Hagedorn temperature $T_H$ (for a discussion of the Horowitz-Polchinski-like solution from the worldsheet in AdS$_3$ see \cite{Nick}). The Hawking-Page temperature $T_{HP}$, where our two winding CFT is lives, lies in between these two critical temperatures (for $k>3$). We expect the two winding CFT to connect to the Horowitz-Polchinski-like solutions around thermal AdS$_3$ or the BTZ black hole as the temperature is varied.\footnote{At the special value of $k=3$ where the dual holographic CFT becomes free, the entire range of temperature (measured in l$_{\text{AdS}}=1$ units) for which the stringy small black hole is expected to exist, $$T_H \geq T \geq \frac{1}{4\pi^2 T_H}$$ collapses to $T = 1/(2\pi)$, the Hawking-Page temperature. Then the radial momentum of both spatial and temporal winding operators are exactly $Q$, and it is natural to conjecture that there is a family of CFTs at this fixed temperature interpolating between thermal AdS$_3$ and BTZ. We thank Yiming Chen for the discussions on this point.}
 
 Instabilities similar to the ones discussed here appear for the black (at temperature $T$) D4 brane wrapped on a circle of radius $R$. The worldvolume theory on the brane (five-dimensional maximally supersymmetric Yang-Mills) has two dimensionless parameters $\lambda /R$, $TR$ ($\lambda$ is the t'Hooft coupling). When $\lambda /R$ is small, the maximally supersymmetric Yang-Mills theory is expected to be confined at sufficiently low temperature (i.e., $TR$ is small).  As the temperature is increased one expects the thermodynamically dominant phase to be composed of a winding condensate on both the Euclidean time and space circle from the point of view of the dual string theory \cite{Aharony:2005ew, Aharony:2006da}.\footnote{This point was also emphasized recently in \cite{Urbach:2023npi}. } When the temperature is higher than a critical value the theory is deconfined and expected to be described by the stringy version of the black D4 brane solution in the bulk. It would be fascinating to understand the role of two winding CFT in this context.

\acknowledgments

    We would like to thank Nick Agia, and Barak Gabai for collaboration on related projects \cite{Nick, Barak} and Yiming Chen,  Nicholas Dorey, Igor Klebanov, Juan Maldacena, Shiraz Minwalla, David Tong, and Xi Yin for insightful conversations. IH is supported by the Harvard Quantum Initiative Fellowship. The work of DLJ is supported in part by DOE grant DE-SC0007870.

\appendix

\bigskip

\providecommand{\href}[2]{#2}\begingroup\raggedright\endgroup

\end{document}